\documentclass[useAMS,usenatbib,fleqn]{mnras}
\usepackage{multirow}
\usepackage{graphicx}
\usepackage{color,soul}
\usepackage{graphicx}
\usepackage{xcolor}
\usepackage{subfig}
\usepackage{epsfig}
\usepackage{epstopdf}
\usepackage{amssymb,amsmath}
\usepackage{aas_macros}
\usepackage[T1]{fontenc}

\title[Polarization ML]{
Using Machine Learning to Link Black Hole Accretion Flows with Spatially Resolved Polarimetric Observables
}

\author[Qiu et al.]{Richard Qiu$^{1,2,3}$, Angelo Ricarte$^{1,4}$, Ramesh Narayan$^{1,4}$, George N.~Wong$^{5,6}$, 
\newauthor
Andrew Chael$^{7}$, and Daniel Palumbo$^{1,4}$
\\
$^1$ Center for Astrophysics | Harvard \& Smithsonian, 60 Garden Street, Cambridge, MA 02138, USA \\
$^2$ Department of Physics, Harvard University, 17 Oxford Street Cambridge, MA 02138, USA \\
$^3$ John A. Paulson School of Engineering and Applied Sciences, Harvard University, 150 Western Ave, Allston, MA 02134, USA \\
$^4$ Black Hole Initiative at Harvard University, 20 Garden Street, Cambridge, MA 02138, USA \\
$^5$ School of Natural Sciences, Institute for Advanced Study, 1 Einstein Drive, Princeton, NJ 08540, USA \\
$^6$ Princeton Gravity Initiative, Princeton University, Princeton, New Jersey 08544, USA \\
$^7$ Princeton Center for Theoretical Science, Princeton University, Jadwin Hall, Princeton, NJ 08544, USA \\
}

\date{\today}

\begin{document}
\pagerange{\pageref{firstpage}--\pageref{lastpage}} \pubyear{2020}
\maketitle

\begin{abstract}
We introduce a new library of 535,194 model images of the supermassive black holes and Event Horizon Telescope (EHT) targets Sgr A* and M87*, computed by performing general relativistic radiative transfer calculations on general relativistic magnetohydrodynamics simulations. Then, to infer underlying black hole and accretion flow parameters (spin, inclination, ion-to-electron temperature ratio, and magnetic field polarity), we train a random forest machine learning model on various hand-picked polarimetric observables computed from each image.  Our random forest is capable of making meaningful predictions of spin, inclination, and the ion-to-electron temperature ratio, but has more difficulty inferring magnetic field polarity.  To disentangle how physical parameters are encoded in different observables, we apply two different metrics to rank the importance of each observable at inferring each physical parameter.  Details of the spatially resolved linear polarization morphology stand out as important discriminators between models.  Bearing in mind the theoretical limitations and incompleteness of our image library, for the real M87* data, our machinery favours high-spin retrograde models with large ion-to-electron temperature ratios.  Due to the time-variable nature of these targets, repeated polarimetric imaging will further improve model inference as the EHT and next-generation (EHT) continue to develop and monitor their targets.  
\end{abstract}

\begin{keywords}
accretion, accretion discs --- black hole physics --- galaxies: individual (M87) --- magnetohydrodynamics (MHD) --- polarization
\end{keywords}

\section{Introduction}
\label{sec:introduction}

Supermassive black holes (SMBHs) are believed to reside at the centres of all or nearly all massive galaxies, some with masses of billions of times that of the sun \citep[e.g.,][]{Kormendy&Richstone1995,Kormendy&Ho2013}.  In the past few years, the Event Horizon Telescope (EHT) collaboration produced the first resolved images of SMBHs, ushering in a new era of resolved SMBH astrophysics \citep{EHTC+2019a,EHTC+2019b,EHTC+2019c,EHTC+2019d,EHTC+2019e,EHTC+2019f,EHTC+2021a,EHTC+2021b,EHTC+2022a,EHTC+2022b,EHTC+2022c,EHTC+2022d,EHTC+2022e,EHTC+2022f}.  So far, published observations include both spatially resolved total intensity and linear polarization maps, while circular polarization, spectral index maps, and rotation measure maps are anticipated.  In the upcoming decade, the next-generation EHT (ngEHT) will improve observing capabilities to include larger bandwidths, additional stations, and additional frequencies. This will enable the production of movies with orders of magnitude of dynamic range that will simultaneously capture disk and jet dynamics \citep{Doeleman+2019,Raymond+2021}.

Spatially resolved polarimetric imaging of these SMBH accretion flows has allowed us to place constraints on aspects of the accretion flow and the spacetime that houses it.  Theoretically interpreting these data has usually involved generating computationally expensive libraries of tens to hundreds of thousands of images originating from general relativistic magnetohydrodynamic (GRMHD) simulations \citep{EHTC+2019e,EHTC+2021b,EHTC+2022e}.  To bridge the gap between theory and observation, observable quantities are computed from each simulated image, which can be compared to the observations.  This methodology has allowed the EHT collaboration to conclude that M87* has dynamically important magnetic fields \citep{EHTC+2021a,EHTC+2021b}, but this can be both cumbersome and inefficient for finding physical trends spanning a multi-dimensional parameter space.  As our observational datasets grow more complex and the theoretical parameter space grows, connecting data and theory will grow increasingly challenging.  

EHT data sit at an intersection between theories of gravity, magnetohydrodynamics, and plasma physics, and thus many theoretical parameters can be jointly constrained.  For each SMBH, one key unknown is its spin, henceforth denoted $a_\bullet \in [-1,1]$\footnote{We use a negative sign to denote a retrograde accretion disk, where the SMBH and accretion disk angular momenta are anti-aligned.}, its dimensionless angular momentum \citep{Kerr1963}.  A SMBH's spin mediates both its accretion and feedback processes:  the radiative efficiency of a thin disk depends on the location of the innermost stable circular orbit \citep[e.g.,][]{Longair2011}, and spin can be extracted to power jets via the magnetic analog of the Penrose process \citep{Penrose&Floyd1971,Blandford&Znajek1977}.  A SMBH's spin also encodes its recent cosmic assembly history \citep[e.g.,][]{Volonteri+2005,Barausse2012}.  Prolonged accretion via a thin disk that maintains its orientation can spin a SMBH up to a maximum value of $a_\bullet=0.998$ \citep{Thorne1974}.  However, thin disk accretion at random orientations will tend to spin a SMBH down on average \citep[e.g.,][]{King+2008}.  At lower Eddington rates, when accretion disks become geometrically thick, even prograde accretion can spin SMBHs down due to the spin extraction required to power jets \citep{Tchekhovskoy+2012,Narayan+2022}.  Spin is also directly impacted by SMBH mergers, which may even dominate low-redshift SMBH growth in the most massive galaxies \citep{Kulier+2015,Ricarte&Natarajan2018,Weinberger+2018,Pacucci&Loeb2020}.  For these reasons, constraining SMBH spins is a key science goal for the ngEHT \citep{Ricarte+2022d}.

In addition to spin, EHT analyses also typically explore different prescriptions for the electron temperature (described in more detail in \autoref{sec:grrt}).  In these rarified accretion flows, the mean free path is much larger than the size scale of the system, causing ions and electrons to separate into a two-temperature plasma \citep{Shapiro+1976,Ichimaru1977,Rees+1982,Narayan&Yi1995b,Yuan2014}.  Significant uncertainties still exist in modelling the heating of electrons, which may be one or two orders of magnitude cooler than the ions in regions where thermal pressure dominates over magnetic pressure \citep{Sadowski+2017,Ryan+2018,Chael+2019,Mizuno+2021}.  Finally, while typically ignored, we also consider the polarity of the magnetic field with respect to the angular momentum of the disk.  As we shall show, this can impart signatures onto both linear and circular polarization \citep[see also][]{Emami+2022}, and may provide insights into how the magnetic field is generated \citep{Contopoulos&Kazanas1998,Contopoulos+2022}.

In \autoref{tab:intro}, we provide a non-exhaustive list of observational measurements accessible to EHT on the left, and important parameters for theoretical models on the right.  We write those which we will consider in this study in black, and additional interesting observational constraints and theoretical explorations in grey.  Additional observations include potentially both resolved and unresolved rotation measure \citep[e.g.,][]{Agol2000,Quataert2000,Marrone2006,Kuo+2014,Goddi+2021,Ricarte+2020} and spectral index \citep[e.g.,][]{Kim+2018,Bower+2019,Ricarte+2023}, which require a significantly more expensive multi-frequency theoretical analysis.  Meanwhile, GRMHD image libraries could be extended to include positrons \citep{Anantua+2020,Emami+2021}, various implementations of non-thermal electrons \citep{Mao+2017,Davelaar+2018,Cruz-Osorio+2022,Fromm+2022,EHTC+2022e}, misaligned disks \citep{Fragile+2007,Liska+2021}, and different elemental abundances \citep{Wong&Gammie2022}, all of which may have significant effects on the observables, but would balloon the dimensionality of a theoretical investigation.  Fortunately, the recent development of machine learning algorithms offers an efficient framework for connecting an increasingly burgeoning theoretical parameter space to an increasingly rich observational dataset.  

\begin{table*}
\begin{tabular}{c|c}
Observable Parameters & Theoretical Parameters \\
\hline
Image Size & Black Hole Spin \\
Image Asymmetry & Inclination \\
Net Linear Polarization Fraction & Magnetic Field Polarity \\
Net Circular Polarization Fraction & Ion-to-Electron Temperature Ratio ($R_\mathrm{high}$) \\
Resolved Linear Polarization Morphology $(\beta_j)$ & \textcolor{gray}{Magnetic Field State} \\
\textcolor{gray}{Resolved Linear Polarization Fraction} & \textcolor{gray}{Positron Fraction} \\ 
\textcolor{gray}{Resolved Circular Polarization Fraction} & \textcolor{gray}{Non-thermal Electron Distribution Slope} \\
\textcolor{gray}{Rotation Measure} & \textcolor{gray}{Disk Tilt} \\
\textcolor{gray}{Spectral Index} & \textcolor{gray}{Hydrogen-to-Helium Ratio} \\
\end{tabular}
\caption{A non-exhaustive list of observables achievable by the EHT or ngEHT as well as theoretical parameters one can constrain using a library of GRMHD simulations.  Parameters considered in this study are shown in black, while additional parameters outside the scope of this work are written in grey.  As both the observational data and our theoretical explorations expand, it is important to devise efficient frameworks to connect observational and theoretical parameters.}
\label{tab:intro}
\end{table*}

In this work, we first develop a novel library of images for Sgr A* and M87* using a suite of MAD General Relativistic Magnetohydrodynamics (GRMHD) simulations first presented in \citet{Narayan+2022}.  These simulations feature densely sampled spin coverage (9 values between $\pm0.9$) and long run times (up to $\approx10^5 \ GM_\bullet/c^3$, where $G$ is the gravitational constant, $M_\bullet$ is the SMBH mass, and $c$ is the speed of light).  Then, we compute a variety of different observable quantities obtainable by EHT studies and apply a machine learning model to identify trends that can allow us to infer quantities such as spin, inclination, and $R_\mathrm{high}$, a parameter associated with electron heating.  Within the context of our incomplete library, we make estimates for M87*'s values of $a_\bullet$, $R_\mathrm{high}$, and magnetic field polarity.  Perhaps more importantly than the predictions themselves, we use the machine learning algorithm to learn how each of the physical parameters is imprinted onto the observational data to provide insights for further development of the EHT and ngEHT.

In \autoref{sec:library}, we describe how our image library is generated and how quantities are computed for comparison with observations.  Then, in \autoref{sec:machine_learning}, we develop a random forest machine learning model to analyse this library.  We use this library to determine how well our model can infer $a_\bullet$, $R_\mathrm{high}$, and magnetic field polarity, apply it to M87* constraints, and test the importance of repeated observations.  Finally, we summarise and discuss the conclusions of this study in \autoref{sec:discussion}.

\section{Image Library}
\label{sec:library}

\begin{figure*}
  \centering
  \includegraphics[width=\textwidth]{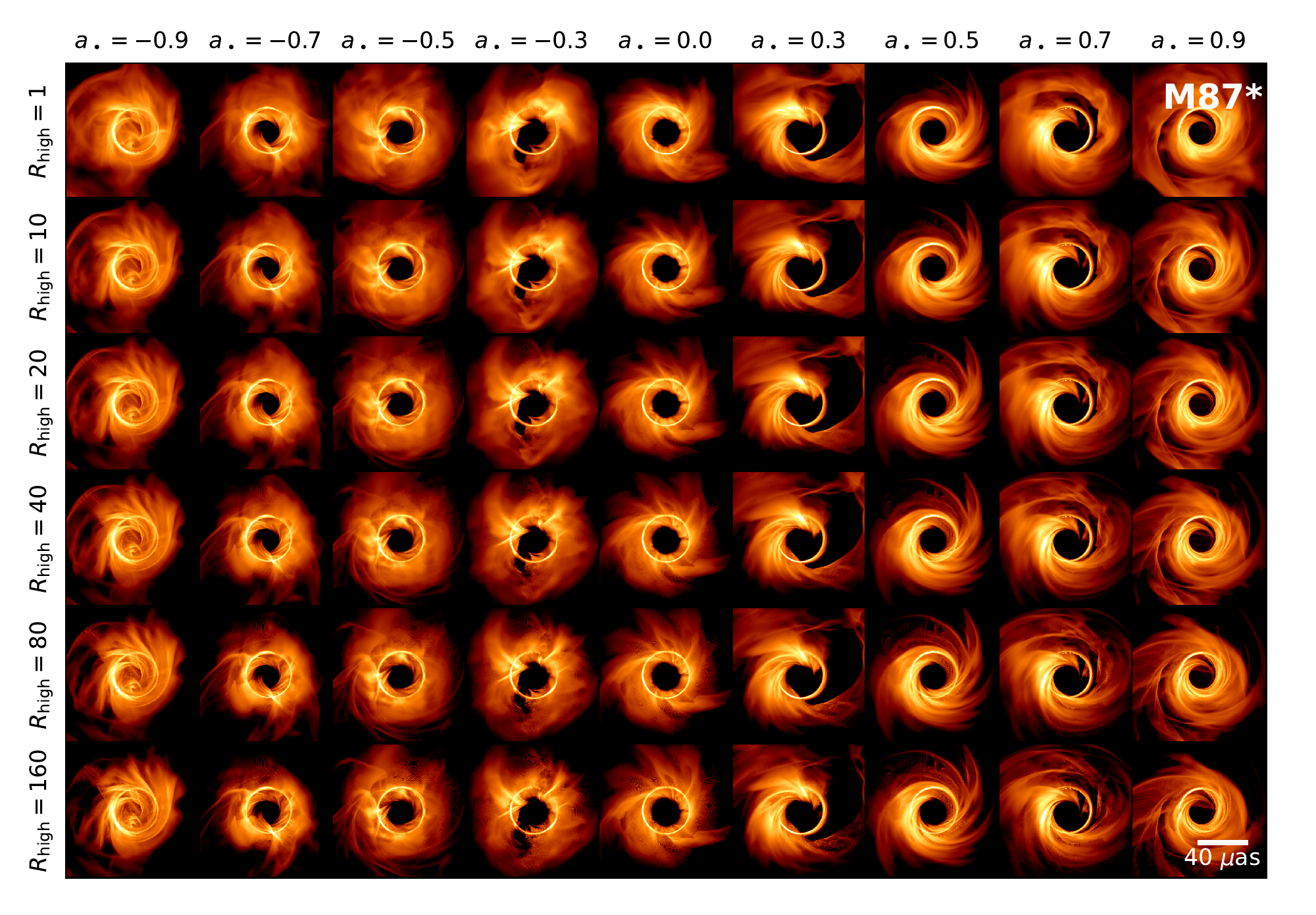}
  \caption{Representative snapshots across the range of spins and $R_\textrm{high}$ in our M87* image library. Prograde and $a_\bullet = 0$ models are traced at inclination $i = 163^\circ$, while retrograde models are traced at inclination $i = 17^\circ$.  Here, the forward-jet is projected straight down, the mean brightness asymmetry appears on the left, and the flow rotates clockwise on the sky.  All models plotted here have magnetic field polarities aligned with the outer disk angular momentum vector.  Some images exhibit cavities from flux eruption events, which occur occasionally in MAD simulations.  Each image is plotted in logarithmic scale with three orders of magnitude dynamic range, normalised individually.  Models with small $|a_\bullet|$ produce more radial inflows than models with large $|a_\bullet|$.  Increasing $R_\mathrm{high}$ typically suppresses disk flux relative to jet flux, but this effect is subtle for MAD models relative to SANEs \citep[e.g.,][]{EHTC+2019e}.
  \label{fig:m87_representative}}
\end{figure*}

\begin{figure*}
  \centering
  \includegraphics[width=\textwidth]{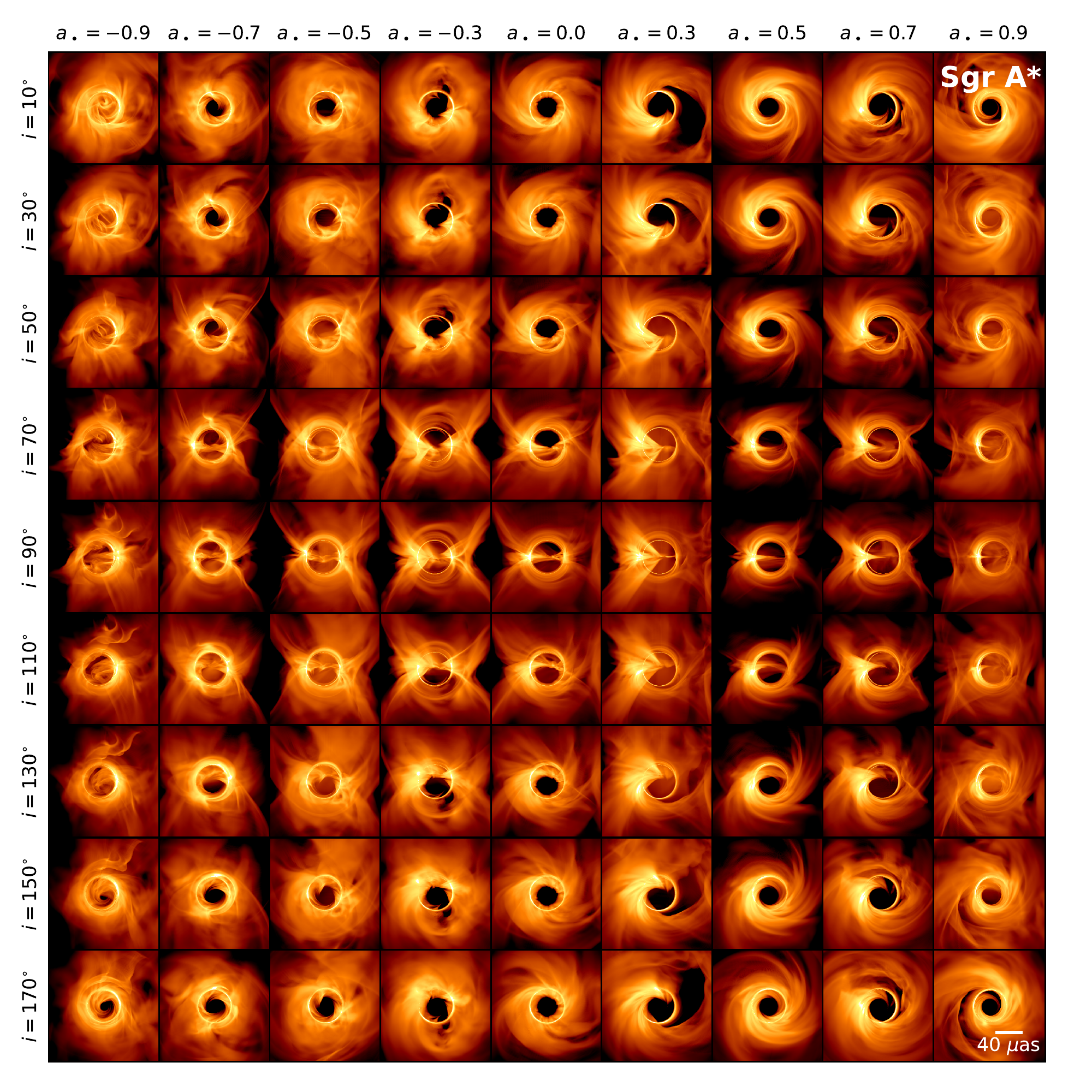}
  \caption{Representative snapshots across the range of spins and inclinations in our Sgr A* image library. Each image is plotted in logarithmic scale with three orders of magnitude dynamic range, normalised individually. Here, we show only $R_\mathrm{high}=20$, but explore different inclinations.  The same GRMHD snapshots are shown as in \autoref{fig:m87_representative}.  Compared to the M87* images, Sgr A* images subtend a larger angle and tend to be more optically thick.
  \label{fig:sgrA_representative}}
\end{figure*}

We consider GRMHD simulations of MAD accretion disks \citep{Narayan+2022} and perform general relativistic radiative transfer (GRRT) with {\sc ipole} to produce images appropriate for Sgr A* and M87* \citep{Moscibrodzka&Gammie2018,Wong+2022}.  We compute observable quantities from these images, then develop a random forest machine learning model that can infer $a_\bullet$, $R_\mathrm{high}$, inclination, and magnetic field polarity from these observables.

\subsection{GRMHD}
\label{sec:grmhd}

The GRMHD simulations used as the starting point of our calculations are presented in \citet{Narayan+2022}.  These simulations were run with the code {\sc KORAL} \citep{Sadowski+2013,Sadowski+2014}, assuming ideal GRMHD, with an adiabatic index of 13/9, a compromise between 5/3 (appropriate for the non-relativistic ions) and 4/3 (appropriate for the relativistic electrons).  We use 9 separate GRMHD simulations, corresponding to different spins $a_\bullet \in \{-0.9, -0.7, -0.5, -0.3, 0.0, 0.3, 0.5, 0.7, 0.9\}$.  These simulations feature a resolution of $288\times 192\times 144$ cells in the $r$, $\theta$, and $\phi$ directions respectively, with an outer radial boundary of $10^5 \ GM_\bullet/c^2$. They are run in a modified version of Kerr-Schild coordinates that concentrate resolution in both the jet and midplane regions.  These simulations are initialized with a standard \citet{Fishbone&Moncrief1976} torus of gas seeded with a weak poloidal magnetic field that extends between $20 \ GM_\bullet/c^2$ and $10^4 \ GM_\bullet/c^2$. Gas is artificially inserted in the zero angular momentum observer (ZAMO) frame to maintain a ceiling on the plasma magnetization of $\sigma \leq 100$.  Each simulation was run for a duration of $\approx 10^5 \ GM_\bullet/c^3$ in search of evolution on long timescales.  As we discuss in Appendix \ref{sec:time_evolution}, we do not find any significant evolution on these timescales apart from an exponential decrease in the accretion rate attributable to the draining and relaxation of the initial torus.  This suggests that simulated images originating from GRMHD are insensitive to the length of the simulation, up to $10^5 \ G M_\bullet/c^3$, as long as the they are sampled beyond the initial relaxation phase.

Each of these simulations quickly reaches the ``Magnetically Arrested Disk'' (MAD) state \citep{Bisnovatyi-Kogan&Ruzmaikin1974,Igumenshchev+2003,Narayan+2003}, which is characterised by relatively ordered and dynamically important magnetic fields with magnetic flux parameter $\phi_\mathrm{BH} \sim 30-50$, where 
\begin{equation}
    \phi_{\rm BH} = \frac{\sqrt{4\pi}}{2\sqrt{\dot{M}_0}} \int_\theta \int_\phi \left|B^r\right|_{r=r_{\rm H}} \;\sqrt{-g}\;\mathrm{d}\theta\; \mathrm{d}\phi
    \label{eq::madparam}
\end{equation}
\noindent is the magnetic flux threading the horizon, normalised by the square-root of the accretion rate \citep{Tchekhovskoy+2011}.  Here, $\dot{M}_0$ is the mass accretion rate through the horizon, located at radius $r_H = 1 + \sqrt{1 - a_\bullet}$.  In this work, we do not consider ``Standard and Normal Evolution'' (SANE) models \citep{Narayan+2012,Sadowski+2013}, which are characterised by weaker and more turbulent magnetic fields.  At present, MAD models are favoured over SANEs by polarized EHT observations of M87* \citep{EHTC+2021b} and more naturally explain flaring activity of Sgr A* \citep{Dexter+2020,Porth+2021,Wielgus+2022}.

\subsection{GRRT}
\label{sec:grrt}

We create images in post-processing using the code {\sc ipole} \citep{Moscibrodzka&Gammie2018}, following standard methodology for library generation \citep{Wong+2022}.  Ideal GRMHD simulations are scale free, and it is at the radiative transfer step where we must specify the mass, distance, accretion rate, viewing angle, and electron distribution function.  The SMBH mass determines the length and time scales for the problem. Meanwhile, the equations of ideal GRMHD are invariant under the transformation
\begin{align}
    \rho &\mapsto \mathcal{M} \rho \\
    u &\mapsto \mathcal{M} u \\
    B &\mapsto \sqrt{\mathcal{M}} B
\end{align}
\noindent where $\rho$ is the mass density, $u$ is the thermal energy density, $B$ is the magnetic field strength, and $\mathcal{M}$ is an arbitrary mass-density scale factor.  We iteratively fit for $\mathcal{M}(t)$ such that the average flux at 230 GHz matches that of the observations.  A novel aspect of our fitting procedure is that rather than fitting for a single scalar, we allow $\mathcal{M}$ to vary with time: we parameterize $\mathcal{M}(t) = \exp(a+bt)$, where $t$ is time in gravitational units, and simultaneously fit for $a$ and $b$.  This parameterization allows us to counteract the artificial decrease in flux that occurs due to the draining and relaxation of the initial torus on long timescales while preserving variability on short timescales, discussed in more detail in Appendix \ref{sec:time_evolution}. We fit $\mathcal{M}$(t) to reproduce an average flux of 0.5 Jy and 2.4 Jy at 230 GHz for our M87* and Sgr A* libraries respectively, consistent with \citet{EHTC+2021b} and \citet{EHTC+2022e}. We imaged our M87* library with a field of view of 160 $\mu$as and an angular resolution of 0.4 $\mu$as. We imaged our Sgr A* library with a field of view of 200 $\mu$as and an angular resolution of 0.5 $\mu$as.

As first defined by \citet{Moscibrodzka+2016}, we set the ratio of ion to electron temperatures using the plasma magnetization using the following prescription: 
\begin{align}
    R \equiv \dfrac{T_{\rm ion}}{T_{\rm electron}} = R_{\rm low} \dfrac{1}{1 + \beta^2} + R_{\rm high} \dfrac{\beta^2}{1 + \beta^2},
\end{align}
where $R_{\rm low}$ and $R_{\rm high}$ are dimensionless scalars and $\beta = P_{\rm gas} / P_{\rm mag}$ is the ratio of the gas to magnetic pressure.  Typically, $\beta$ is smaller in jet/funnel regions of a simulation compared to mid-plane regions.  Consequently, increasing $R_\mathrm{high}$ tends to move emission from the midplane to the jet/funnel, although the effect is much more dramatic for SANE simulations than for MADs \citep{EHTC+2019e}.  By cooling the midplane, increasing $R_\mathrm{high}$ requires larger values of $\mathcal{M}$ to match the 230 GHz flux, and both higher mass densities and lower temperatures result in larger Faraday rotation depths.  During ray-tracing we zero the radiative transfer coefficients in any regions where the plasma magnetization $\sigma>1$, where numerical floors may artifically inject material.

\begin{table*}
\begin{tabular}{c|c|c}
Parameter & M87* Library Values & Sgr A* Library Values \\
\hline
$a_\bullet$ & $0$, $\pm 0.3$, $\pm 0.5$, $\pm 0.7$, $\pm 0.9$ & $0$, $\pm 0.3$, $\pm 0.5$, $\pm 0.7$, $\pm 0.9$
\\
$R_\mathrm{high}$ & $1$, $10$, $20$, $40$, $80$, $160$ & $1$, $10$, $20$, $40$, $80$, $160$
\\
$R_\mathrm{low}$ & 1 & 1
\\
\multirow{2}{*}{$i$} & $163^\circ$ for $a_\bullet \geq 0$ & \multirow{2}{*}{$10^\circ, 30^\circ, 50^\circ, 70^\circ, 90^\circ, 110^\circ, 130^\circ, 150^\circ, 170^\circ$}
\\
& $17^\circ$ for $a_\bullet < 0$ &
\\
$B$-field & aligned, anti-aligned & aligned 
\end{tabular}
\caption{A summary of the GRMHD and imaging parameters spanned by our M87* and Sgr A* libraries.  Compared to the \citet{EHTC+2021b} library for M87*, we consider more spins, slightly more values of $R_\mathrm{high}$, and both polarities of the magnetic field, but only one value of $R_\mathrm{low}$.}
\label{tab:library_params}
\end{table*}

From the 9 GRMHD simulations, we generate two image libraries, one for M87* and one for Sgr A*. Both libraries span 6 values of $R_\textrm{high} = \{1, 10, 20, 40, 80, 160\}$. Although \citet{EHTC+2021b} considered both $R_\mathrm{low}=1$ and $R_\mathrm{low}=10$, we limit our study to only $R_\mathrm{low}=1$. We briefly test the effect of varying $R_\mathrm{low}$ in Appendix \ref{sec:r_low_test} for a subset of our M87* models.  Sgr A*'s inclination is presently not directly constrained, and thus its library spans 9 inclinations with $i = \{10^\circ, 30^\circ, 50^\circ, 70^\circ, 90^\circ, 110^\circ, 130^\circ, 150^\circ, 170^\circ\}$. Meanwhile, M87*'s large-scale jet and the orientation of its brightness asymmetry directly constrain its inclination.  Prograde and zero spin simulations in the M87* library are imaged at $i = 163^\circ$ and retrograde spin simulations at $i = 17^\circ$ in order to preserve the observed orientation of the brightness asymmetry. Each parameter set is imaged at 901 simulation snapshots from 10,000 $GM_\bullet/c^3$ to 100,000 $GM_\bullet/c^3$ uniformly spaced 100 $GM_\bullet/c^3$ apart. It has recently been appreciated that the poloidal magnetic field direction matters not only for circular polarization, but also for the overall twisty morphology of linear polarization ticks \citep{Emami+2022}.  For the M87* library, we generate images with both polarities of the magnetic field, either aligned with the disk angular momentum vector or anti-aligned (henceforth simply ``aligned'' or ``anti-aligned'').  We only compute the aligned case for Sgr A* simply due to the increased computational expense, as this library contains far more inclinations.

In total, the M87* and Sgr A* libraries have 97,308 and 437,886 images, respectively. Figures \ref{fig:m87_representative} and \ref{fig:sgrA_representative} display representative images from the M87* and Sgr A* libraries, each presented in logarithmic scale. The $R_\textrm{high} = 20$ images in the Sgr A* library were included in the analysis performed in \citet{Georgiev+2022} (as image set C) and \citet{EHTC+2022e}, where they were found to be broadly consistent with other image libraries. In those papers, only our library employed an exponential fit for $\mathcal{M}$ due to the uniquely long GRMHD timescale probed. A smaller set of M87* images created using this methodology were also used in \citet{Ricarte+2022b} to study signatures of retrograde accretion flows.

In \autoref{fig:m87_representative}, we plot models with different values of $a_\bullet$ and $R_\mathrm{high}$.  All images are optically thin enough to feature a clear photon ring, composed of light making multiple orbits around the BH, and an inner shadow.  In logarithmic scale, one can discern thin streams of gas, which sometimes turn around in the retrograde case as explored in \citet{Ricarte+2022b}.  The effect of varying $R_\mathrm{high}$ in our MAD models is quite subtle in total intensity, but as discussed in future sections, is more noticeable in polarization.  In \autoref{fig:sgrA_representative}, we fix $R_\mathrm{high}=20$, but display different inclinations.  Compared to M87*, these images are more optically thick \citep[see also][]{Ricarte+2023}, but still optically thin enough to see the photon ring and inner shadow at all inclinations.  At large inclinations, emission from the jet funnel is more obvious and separable from the disk.  Some snapshots such as $a_\bullet=0.3$ feature large cavities due to ``flux eruption events,'' behaviour characteristic of the MAD state that is implicated in polarized flares \citep[e.g.,][]{Tchekhovskoy+2011,Dexter+2020,Porth+2021,Chatterjee&Narayan2022,Ripperda+2022,Gelles+2022,Wielgus+2022}.  

\subsection{Observable Image Quantities}
\label{sec:observables}

As described in \autoref{sec:grrt}, our dataset for analysis consists of 535,194 images spanning 4 free parameters, each of which contains a wealth of information.  Despite this large number of images, each of these parameters is sampled somewhat coarsely, a problem that will inevitably grow more intractable as additional parameters are considered.  This motivates a machine learning model to efficiently identify trends and make inferences about models.  

Since we still do not know all of the characteristics of polarized images that may carry useful information, several recent studies have trained deep convolutional neural networks on raw image data directly.  \citet{vanderGucht+2020} trained a Bayesian convolutional neural networks to predict spin, inclination, $R_\textrm{high}$, $M$, $\dot{M}$, and position angle from a library of M87* images. Considering only SANE models without polarization information, they found that at the $\sim 20 \mu$as resolution of the EHT, their models could accurately recover $M$ and $\dot{M}$. \citet{Lin+2021} fine-tuned pretrained convolutional neural networks to predict accretion state (MAD or SANE) and spin, both of which they showed can be recovered with high accuracy. However, unlike \cite{vanderGucht+2020}, they did not consider EHT resolution limitations. As a result, they found that their model honed in on low level surface brightness features which are not resolvable by the EHT at current resolutions. And like \cite{vanderGucht+2020}, they only consider the total intensity image, while EHT observations include polarimetric data which have been shown to be important for discriminating between models \citep{EHTC+2021b}. 

Our approach differs from these previous works in two key ways: (i) our analysis include polarimetric information, and (ii) rather than working with raw image data, we first compute pre-processed observables that the machine learning model receives as input.  This approach has several advantages over a neural network approach.  First, neural networks are notoriously difficult to interpret, and their extreme model capacity may cause them to use untrustworthy aspects of images for model discrimination, including numerical artifacts or image details that are impossible to observe in practice.  Our methodology ensures that the model only sees information that we believe is both trustworthy and observable.  Another significant advantage of this methodology is that we can also directly rank the importance of each observable in a straightforward and intuitive manner.  On the other hand, the clear downside of our approach is that we may not include important observables that we have not identified.

Based on previous EHT-related studies, we have identified the following image integrated quantities as important, each computed after first blurring images with a 20 $\mu$as Gaussian beam:
\begin{enumerate}
    \item Total unresolved linear polarization fraction 
        \begin{equation}
            |m|_\mathrm{net} = \frac{\sqrt{\left(\sum_j \mathcal{Q}_j\right)^2 + \left(\sum_j \mathcal{U}_j\right)^2}}{\sum_j \mathcal{I}_j}
        \end{equation}
    \item Total unresolved signed circular polarization fraction
        \begin{equation}
            v_\mathrm{net} = \frac{\sum_j \mathcal{V}_j}{\sum_j \mathcal{I}_j}
        \end{equation}
    \item The first few modes (amplitudes and phases) of radially integrated Fourier decompositions of the azimuthal linear polarization pattern:  $\beta_j$, $j \in \{0, 1, 2, 3 \}$ \citep{Palumbo+2020}
    \item Image asymmetry, $A$ \citep{Medeiros+2022}
    \item Image size (the average of its major and minor axes of the total intensity image)
\end{enumerate}

Here, $\{\mathcal{I}_j,\mathcal{Q}_j,\mathcal{U}_j,\mathcal{V}_j\}$ refer to the Stokes parameters computed in pixel $j$.  We note that the unresolved electric vector position angle (EVPA), defined east {\it relative to the SMBH spin axis} as $\chi = \frac{1}{2} \mathrm{arctan} \frac{\sum_j U_j}{\sum_j Q_j}$, is related to $\angle\beta_0$ by $\chi = \frac{1}{2}\angle\beta_0$. Finally, we note that only $\angle\beta_2$ is invariant under rotation of the image. For M87*, we know the orientation of the jet on the sky observationally, and thus have an absolute reference for $\angle\beta_0$, $\angle\beta_1$, $\angle\beta_3$. However, for Sgr A* (or for a generic low-luminosity AGN ngEHT may observe in the future), we have no such prior knowledge at present. Thus, we use $\angle\beta_1$ as a reference angle and instead use in our machine learning analysis the rotation invariant quantities $\angle\beta_0^\textrm{(Sgr A*)} = \angle\beta_0 - 2 \angle\beta_1$ and $\angle\beta_3^\textrm{(Sgr A*)} = \angle\beta_3 + \angle\beta_1$ and omit $\angle\beta_1$ from our set of observables for Sgr A*. Similarly, \cite{Medeiros+2022} calculate the image asymmetry across the midplane of the image, defined as orthogonal relative to the SMBH spin axis, which they show robustly maximizes the asymmetry across angles on the image. Thus, as we may not know the SMBH spin axis a priori, we calculate the rotationally invariant asymmetry $A^\textrm{(Sgr A*)} = \max_\theta A(\theta)$.

The unresolved linear polarization fraction encodes information about the magnetic field geometry. In particular, it indirectly contains information about symmetries in the geometry (e.g. due to cancellations due to symmetries that flip the angle of polarization) and scrambling due to Faraday rotation. 

The unresolved circular polarization fraction is affected by (i) direct circularly polarized synchrotron emission, (ii) Faraday conversion that exchanges linear and circular polarization, (iii) Faraday rotation that can rotate or potentially scramble linear polarization that can be converted into circular \citep{Wardle&Homan2003}.  It is sensitive to both the direction of the magnetic field and its overall geometry \citep{Ricarte+2021}.  \citet{EHTC+2021b} found that some SANE models had too much circular polarization due to large Faraday conversion depths.  As we shall show, the unresolved circular polarization fraction is one of the most important parameters for inferring the magnetic field's polarity, since both the circular polarization emission coefficient and Faraday rotation coefficient switch sign upon a flip of the magnetic field, but the Faraday conversion coefficient does not \citep[e.g.,][]{Pandya+2016,Dexter2016}.

The resolved linear polarization structure of an image encodes the structure of the near-horizon magnetic field. In particular, the direction of polarized emission is perpendicular to the magnetic field and wavevector. The structure of linear polarization in the resolved image is complicated by Faraday rotation and relativistic effects, but still encodes useful information about the magnetic field structure. As such, we include the argument and magnitude of the 0th through 3rd radially integrated Fourier modes of linear polarization ($\beta_0$, $\beta_1$, $\beta_2$, and $\beta_3$). Notably, \cite{Palumbo+2020} found that ``twisty'' linear polarization structure represented by  $\angle\beta_2$ is highly discriminating for spin within a library of M87* images. Likewise, \cite{EHTC+2021b} found that among all constraints, limits on $\beta_2$ are the most discriminating among various models.

Beyond polarized image observables considered by \cite{EHTC+2021b}, we also consider observables which characterise the resolved total intensity image: brightness asymmetry $A$ and the mean second image moment. The brightness asymmetry in a resolved image is caused by Doppler beaming and boosting which in turn capture information about the near-horizon velocity of the accretion flow. As these are influenced by the viewer's observing angle and spin of the black hole, brightness asymmetry may be a useful discriminant for these BH properties \citep{Medeiros+2022}. Similarly, the image size is influenced by both Doppler effects as well as directly by observing inclination. Additionally, different electron heating prescriptions influence what regions of the accretion flow and jet emit, so the image size may also help constrain the ion-to-electron temperature ratio. 

\subsection{Distributions of Observables}
\label{sec:obs_distributions}

\begin{figure*}
  \centering
  \includegraphics[width=\textwidth]{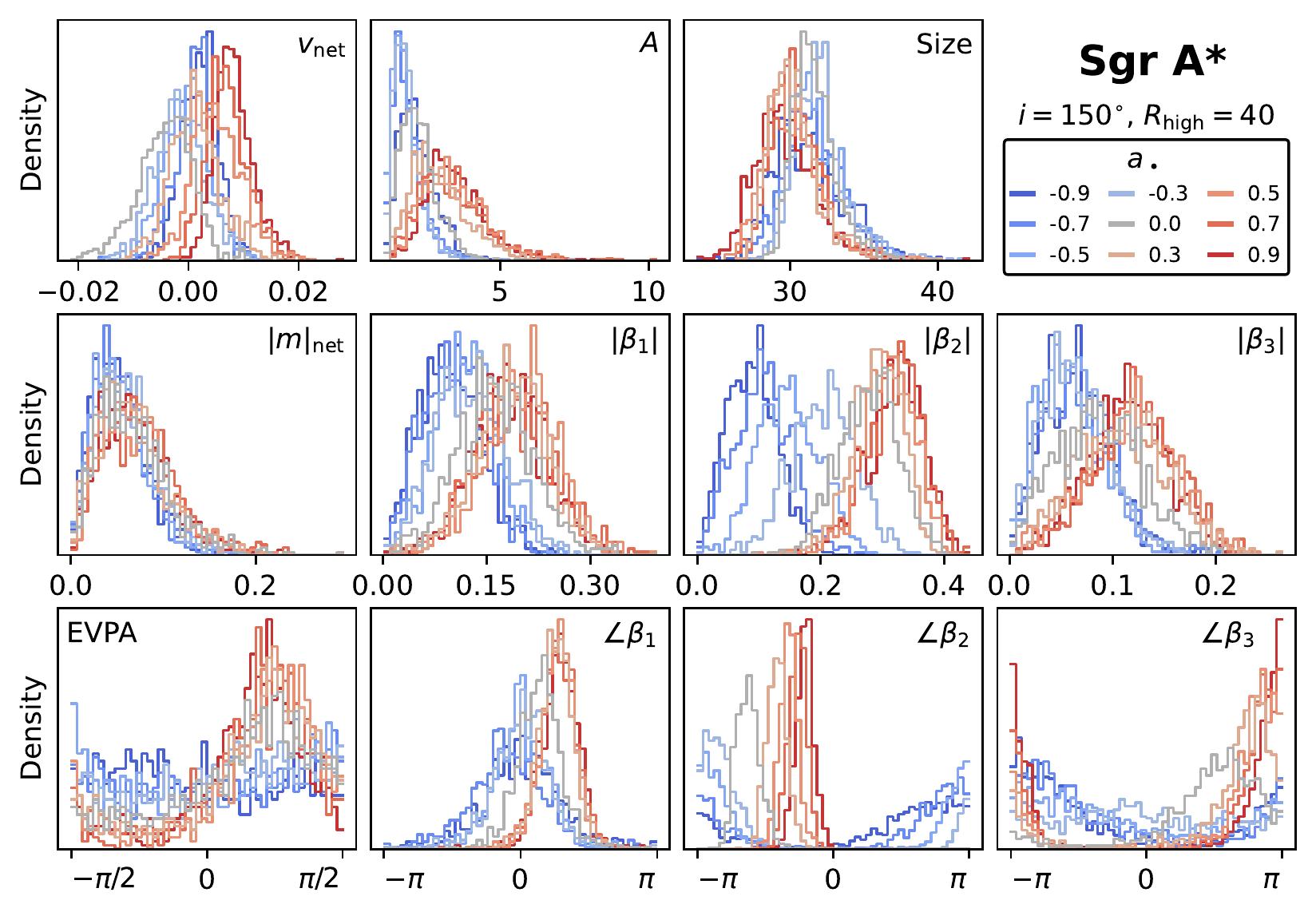}
  \caption{Distributions of observables among our Sgr A* models with $i = 150^\circ$ and $R_\textrm{high} = 40$ as a function of spin.  At this inclination and $R_\textrm{high}$, both the magnitude and phase of $\beta_2$ trend strongly with spin.  Note the strong evolution of $\beta_2$, which encodes the twisty morphology of linear polarization ticks.
  \label{fig:histograms_sgrA_spin}}
\end{figure*}

\begin{figure*}
  \centering
  \includegraphics[width=\textwidth]{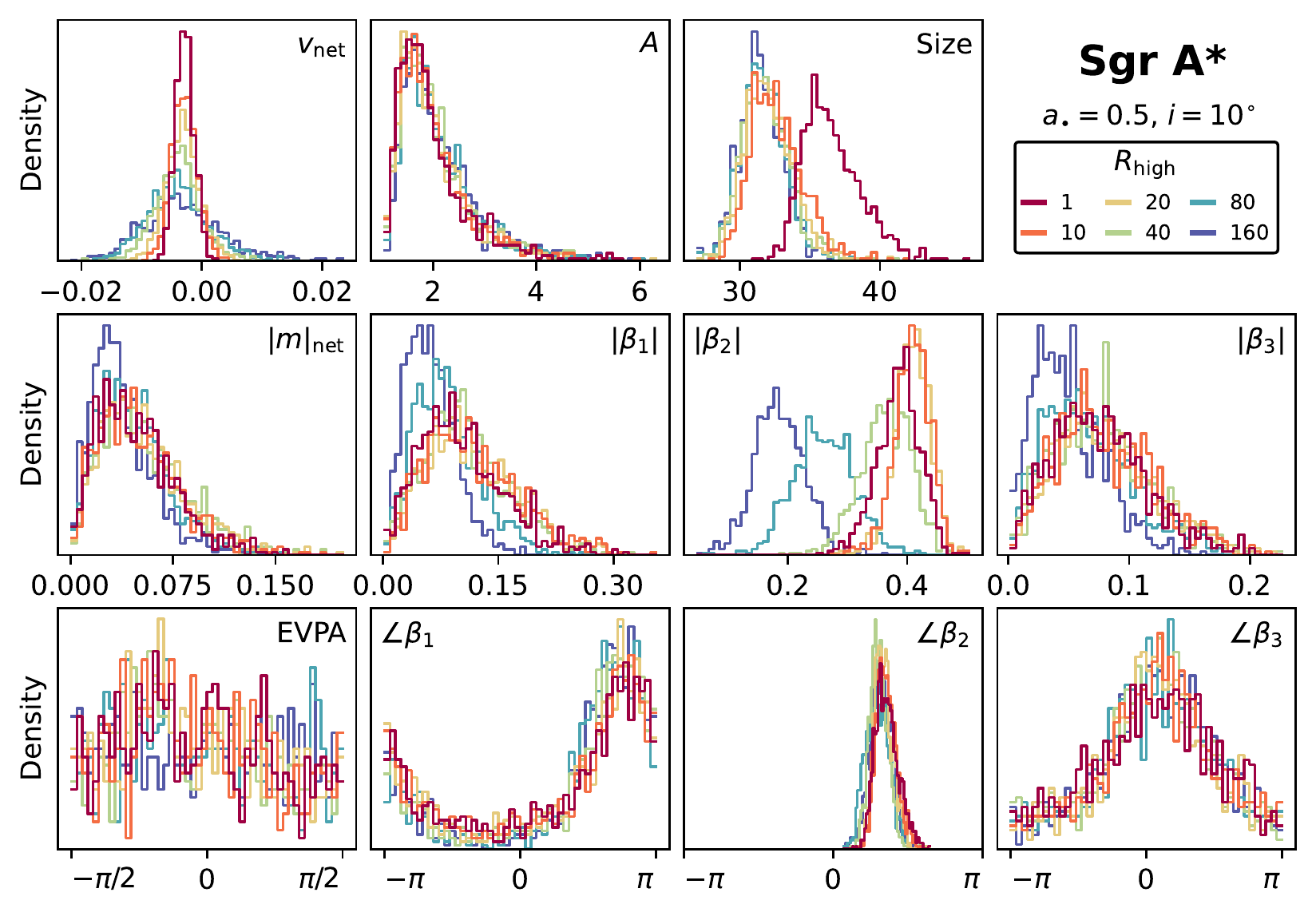}
  \caption{Distributions of observables among our Sgr A* models with $a_\bullet = 0.5$ and aligned magnetic field as a function of $R_\mathrm{high}$.  Larger values of $R_\mathrm{high}$ cool the mid-plane by construction.  Models with larger $R_\mathrm{high}$ have more jet emission and stronger Faraday effects.  Models with larger $R_\mathrm{high}$ therefore have more scrambled linear polarization patterns (weaker $\beta$ mode amplitudes).  Increased Faraday rotation and a shifted emission region also impart a subtle shift in $\beta$ mode arguments as $R_\mathrm{high}$ increases.
  \label{fig:histograms_sgrA_rhigh}}
\end{figure*}

\begin{figure*}
  \centering
  \includegraphics[width=\textwidth]{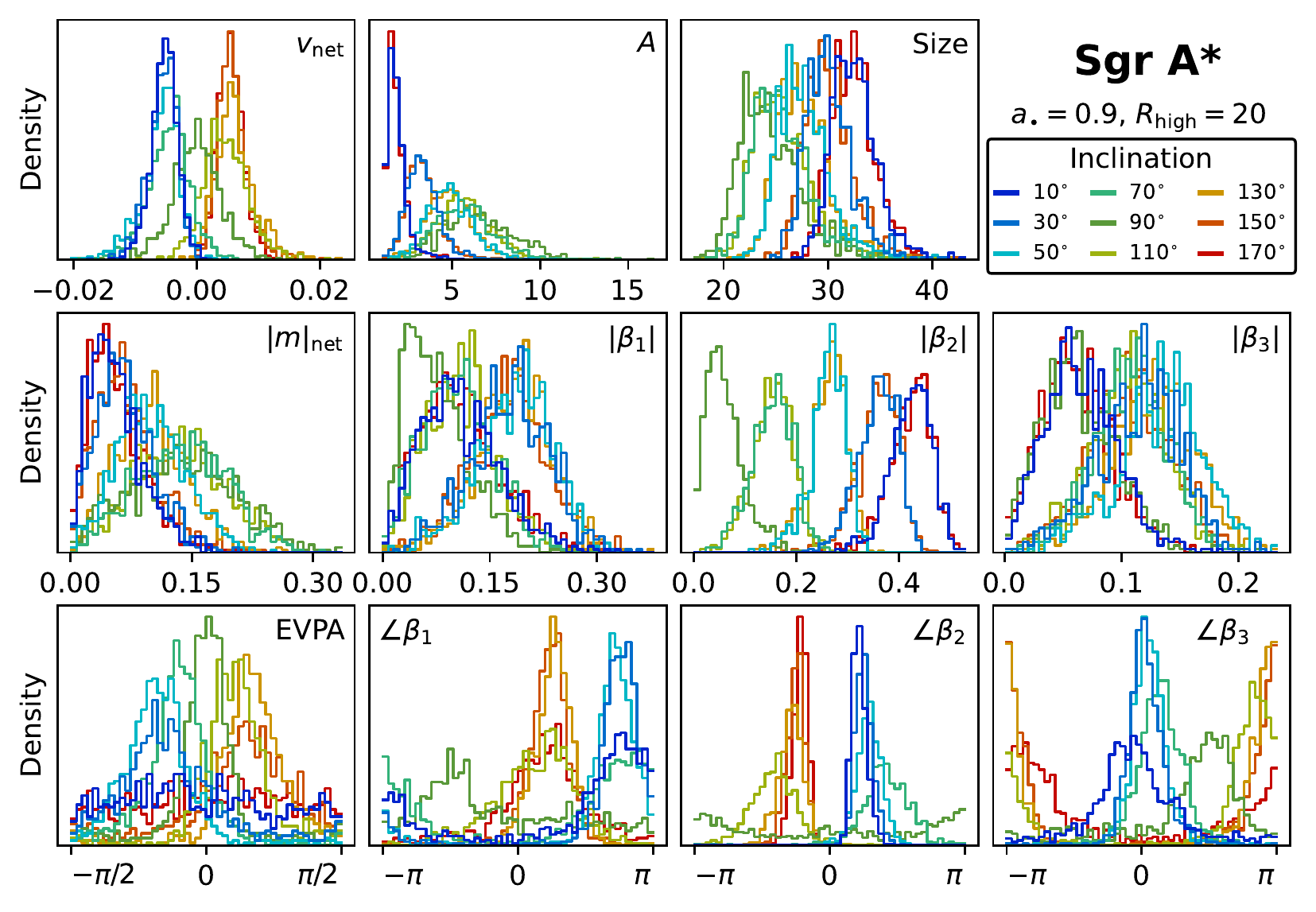}
  \caption{Distributions of observables among our Sgr A* models with $a_\bullet = 0.9$ and $R_\textrm{high} = 20$ as a function of inclination.  For these models, flipping the viewing angle negates the distribution of $v_\mathrm{net}$.  (We caution that this is not a generic result. Although Faraday rotation and intrinsic emission of circular polarization flip sign with flipped orientation, the Faraday conversion coefficient does not \citep{Ricarte+2021}.)  Doppler beaming concentrates emission into a smaller area at large inclinations \citep[see also][]{EHTC+2022e}.  Although one might expect increased Faraday rotation from the mid-plane projected into our line-of-sight to decrease $|m|_\mathrm{net}$ as inclination increases, it instead increases because image symmetry causes cancellation at face-on inclinations.
  \label{fig:histograms_sgrA_incs}}
\end{figure*}

\begin{figure*}
  \centering
  \includegraphics[width=\textwidth]{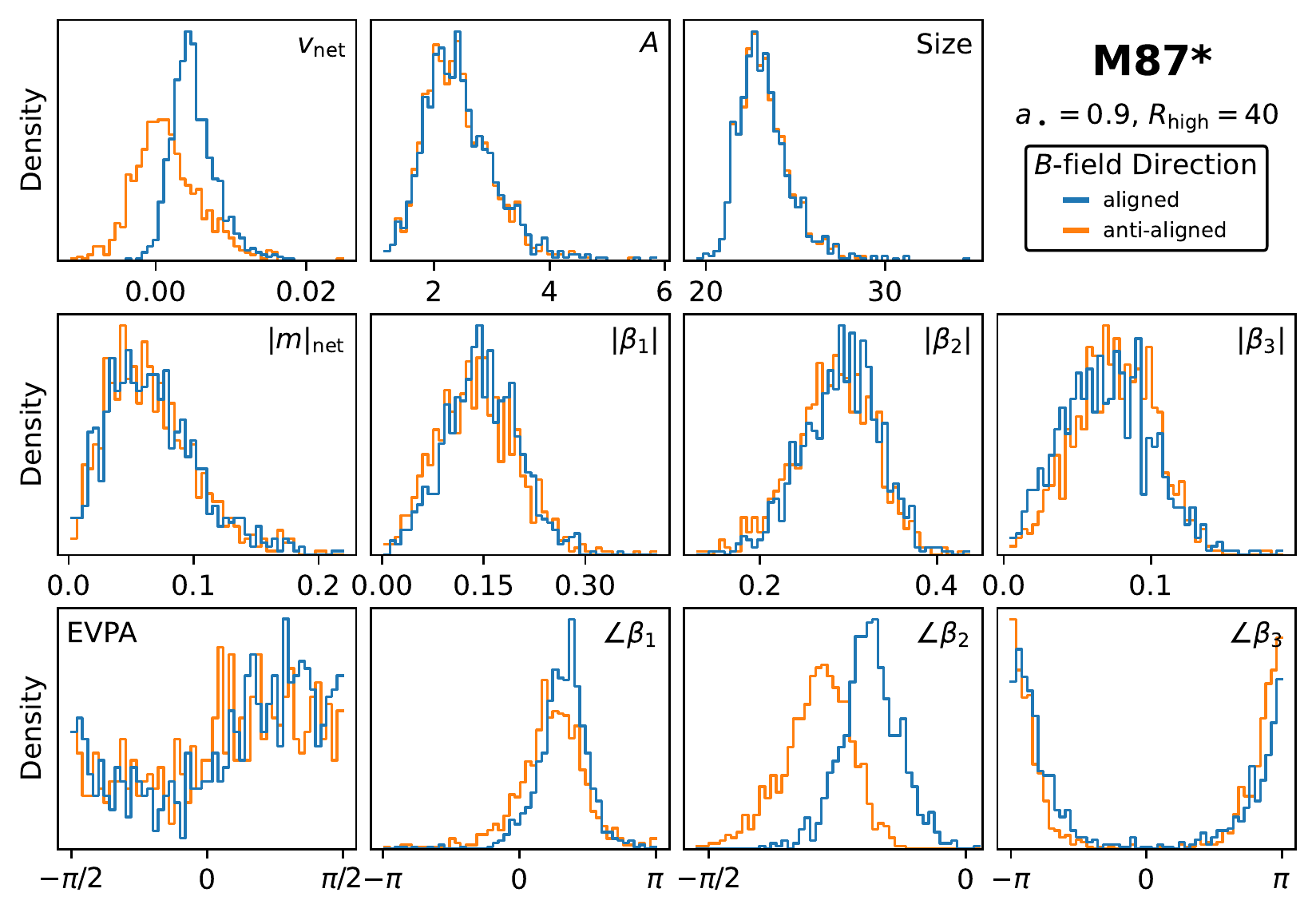}
  \caption{Distributions of observables among our M87* models with $a_\bullet = 0.9$ and $R_\textrm{high} = 40$ as a function of magnetic field polarity with respect to the disk angular momentum vector, which is either aligned or anti-aligned.  While most distributions are identical, there are noticeable shifts in $v_\mathrm{net}$ and $\angle \beta_2$.  Note that the distribution of $v_\mathrm{net}$ does not simply flip across $v_\mathrm{net}=0$.  This is because flipping the magnetic field direction flips the Faraday rotation and circular polarization emissivity coefficients, but not the Faraday conversion coefficient \citep[see][for more detail]{Ricarte+2021}.  Meanwhile, the noticeable shift in $\beta_2$ can be explained entirely by Faraday rotation.  A magnetic field pointing towards us rotates linear polarization ticks counter-clockwise, and vice-versa for a field pointing away from us. 
  \label{fig:histograms_m87_bfield}}
\end{figure*}

For each image, we compute the image-integrated observable quantities outlined in section \ref{sec:observables}. Before developing our machine learning model, we first briefly explore distributions of observables among a few slices of the parameter space to gain an intuition of relevant trends.

\subsubsection{Varying Spin}
\label{sec:trends_spin}

In \autoref{fig:histograms_sgrA_spin}, we first explore trends in our observables as a function of spin.  Here, we include only a subset of our Sgr A* models, where we have fixed $i=150^\circ$ and $R_\textrm{high} = 40$.  Spin affects the dynamics of the inflowing gas, which can be reflected in the polarization structure and Doppler boosting. 

Spin is imprinted onto the total intensity features that we have selected somewhat weakly.  First, larger prograde spins result in smaller image sizes.  These models have the greatest Doppler beaming, focusing the emission into a smaller region.  The same physics results in larger brightness asymmetries for the prograde models compared to the retrogrades, consistent with \citet{Medeiros+2022}.  These effects are more dramatic for more edge-on inclinations.

Clearer trends are imprinted onto the polarization structure, especially the $\beta_2$ coefficient, as first noticed in \citet{Palumbo+2020}.  Recall that $\beta_2$ describes the rotationally symmetric mode of a full Fourier decomposition of the emission.  This mode is believed to trace the underlying magnetic field structure, which in turn is affected by frame dragging \citep{Emami+2022}.  In retrograde models, which have more complex emission morphologies the polarization structure is inherently messier and thus $|\beta_j|$ is lower.  For a fixed inclination, note that $\angle \beta_2$ flips sign for retrogrades compared to progrades, since the spin axis is pointed in the opposite direction with respect to the observer.  Meanwhile, the circular polarization distributions exhibit complex trends that we caution are sensitive to all of the other parameters: inclination, $R_\mathrm{high}$, and magnetic field polarity.  This is because intrinsic emission and Faraday rotation switch sign as the magnetic field polarity switches sign, but Faraday conversion does not.  The relative importances of intrinsic emission and Faraday conversion also depend on inclination and $R_\mathrm{high}$.  

\subsubsection{Varying $R_\textrm{high}$}
\label{sec:trends_rhigh}

In \autoref{fig:histograms_sgrA_rhigh}, we now vary $R_\mathrm{high}$ for a subset of the Sgr A* models, keeping fixed a moderate prograde spin $a_\bullet = 0.5$ and observing inclination $i=10^\circ$.  Increasing $R_\mathrm{high}$ tends to shift emission from the mid-plane to the jet funnel, although the effect is not as pronounced for MADs as it is for SANEs \citep{EHTC+2019e}.

In total intensity, for this face-on model, $R_\mathrm{high}=1$ models have the largest sizes, since they light up disk material.  Material in the funnel is always projected to relatively small image radius for this inclination.  Meanwhile, in polarization, first notice that $|\beta_j|$ tends to decreases with increasing $R_\mathrm{high}$.  Models with larger $R_\mathrm{high}$ have larger Faraday depths, as explored in \citet{EHTC+2021b}.  Colder electrons are more efficient at Faraday rotation, and these models are also normalised to have higher density.  Since $|\beta_j|$ is correlated with the image-averaged polarization, more Faraday rotation results in smaller $|\beta_j|$.  We also notice a very small shift in $\angle \beta_2$ as a function of $R_\mathrm{high}$ due to different amounts of Faraday rotation, first pointed out in \citet{Emami+2022}.  As we shall show, since $\angle \beta_2$ is not {\it too} sensitive to $R_\mathrm{high}$ at low inclination, it is a good predictor of spin.  Regarding circular polarization, greater values of $R_\mathrm{high}$ lead to wider distributions of $v_\mathrm{net}$ for this model, though we suspect this will depend on the detailed interplay of Faraday conversion and intrinsic emission in any particular model.  At higher inclinations, Faraday depth becomes more important, since emission now must pass through colder disk material on the way to the camera.  We find that for larger viewing angles, increasing $R_\mathrm{high}$ has more dramatic effects decreasing $|\beta_j|$ and scrambling their phases.

\subsubsection{Varying Inclination}
\label{sec:trends_inc}

In \autoref{fig:histograms_sgrA_incs}, we explore how our observables vary as a function of inclination for the subset of our Sgr A* images with $a_\bullet=0.9$ and $R_\mathrm{high}=20$.  Important physical effects that vary with inclination include the strength of Doppler beaming, and the evolution of the radiative transfer coefficients that are sensitive to magnetic field direction with respect to the photon wavevector, especially $j_V$ and $\rho_V$, which describe circularly polarized emission and Faraday rotation.

In total intensity, we observe that edge-on inclinations lead to smaller and more asymmetric images.  Doppler boosting is most effective at edge-on inclinations, resulting in emission concentrated in a relatively small area on one side of the image.  Note that the trend may be reversed for different models dominated by jet emission such as large $R_\mathrm{high}$ SANEs, which can be projected to a larger extent at edge-on inclinations.

There are several interesting trends in polarized quantities.  Counterintuitively, $|m|_\mathrm{net}$ is minimised for face-on inclinations at the same time that $|\beta_2|$ is maximised.  This is because this model is very symmetric in linear polarization when viewed face-on, leading to cancellation when the linear polarization is summed \citep[e.g.,][]{Ricarte+2022c}.  Unsurprisingly, the rotationally symmetric $\beta_2$ mode is strongest when viewing the disk face-on.  Also, viewing the system from the opposite side flips the handedness of the linear polarization ticks, which corresponds to a flip in the sign of $\angle \beta_2$.  Interestingly, the amplitudes $|\beta_{1,3}|$ are highest at intermediate inclinations.  As the system is tilted away from face-on, the power in the rotationally symmetric mode spills onto nearby modes.  In the distributions of $v_\mathrm{net}$, we see some moderate shifts as a function of inclination and a sign flip when flipping the viewing angle. This is because the sign of circular polarization from intrinsic emission and Faraday rotation flip with the viewing angle. However, we caution that flipping the viewing angle does not always negate the circular polarized image \citep{Ricarte+2021}.

\subsubsection{Trends as a Function of Magnetic Field Polarity}
\label{sec:trends_bfield}

Finally, in \autoref{fig:histograms_m87_bfield}, we explore the effect of flipping the magnetic field polarity in the subset of our M87* models with $a_\bullet = 0.9$ ($i = 163^\circ$) and $R_\textrm{high} = 40$.  Both the circular polarization emission coefficient $j_V$ and the Faraday rotation coefficient $\rho_V$ flip sign upon flipping the magnetic field direction, leading to cascading effects.

Although most quantities are insensitive to the magnetic field polarity, we observe interesting shifts in $v_\mathrm{net}$ and $\angle \beta_2$.  Interestingly, the distribution of $v_\mathrm{net}$ does not simply reflect across $v_\mathrm{net}=0$ upon flipping the magnetic field direction, but the nature of the distribution changes entirely.  In this model, an aligned magnetic field polarity with the disk angular momentum results in $v_\mathrm{net}>0$ almost always, whereas the anti-aligned case yields a more symmetric distribution.  As discussed in \citet{Ricarte+2021}, flipping the magnetic field flips the circular polarization emission coefficient $j_V$ and the Faraday rotation coefficient $\rho_V$, but not the Faraday conversion coefficient $\rho_Q$.  Depending on the model and magnetic field polarity, the circular polarization from intrinsic emission and Faraday conversion may add or cancel.  This will be explored in much more detail using a library of simulations in Joshi et al.~(in prep.).

A systematic shift occurs in $\angle\beta_2$ upon flipping the magnetic field, also explored in \citet{Emami+2022}.  This occurs because the magnetic field polarity imparts a systematic shift in all of the linear polarization ticks due to Faraday rotation, which switches direction upon flipping the field.

\section{Inferring Model Parameters with Machine Learning}
\label{sec:machine_learning}

Using the quantities described in \ref{sec:observables} as predictors, we perform regression on spin, inclination, $R_\textrm{high}$, and magnetic field polarity by building a random forest machine learning model. Generically, we expect the functional relationship between these quantities and spin to be complicated. Random forests are particularly well-suited for this task due to their ability to characterise arbitrarily complicated relationships (given sufficiently large data), relative robustness against overfitting, and relative lack of hyperparameters (and therefore little hand-tuning needed to achieve strong predictive performance) \citep{Breiman2001}. 
We give a brief overview of random forests and their training algorithms here, but refer readers to \cite{Breiman2001, Hastie+2009, Mehta+2019} for a more detailed explanation. 

\subsection{Random Forests}
\label{sec:rfs}

\begin{figure*}
  \centering
  \includegraphics[width=\textwidth]{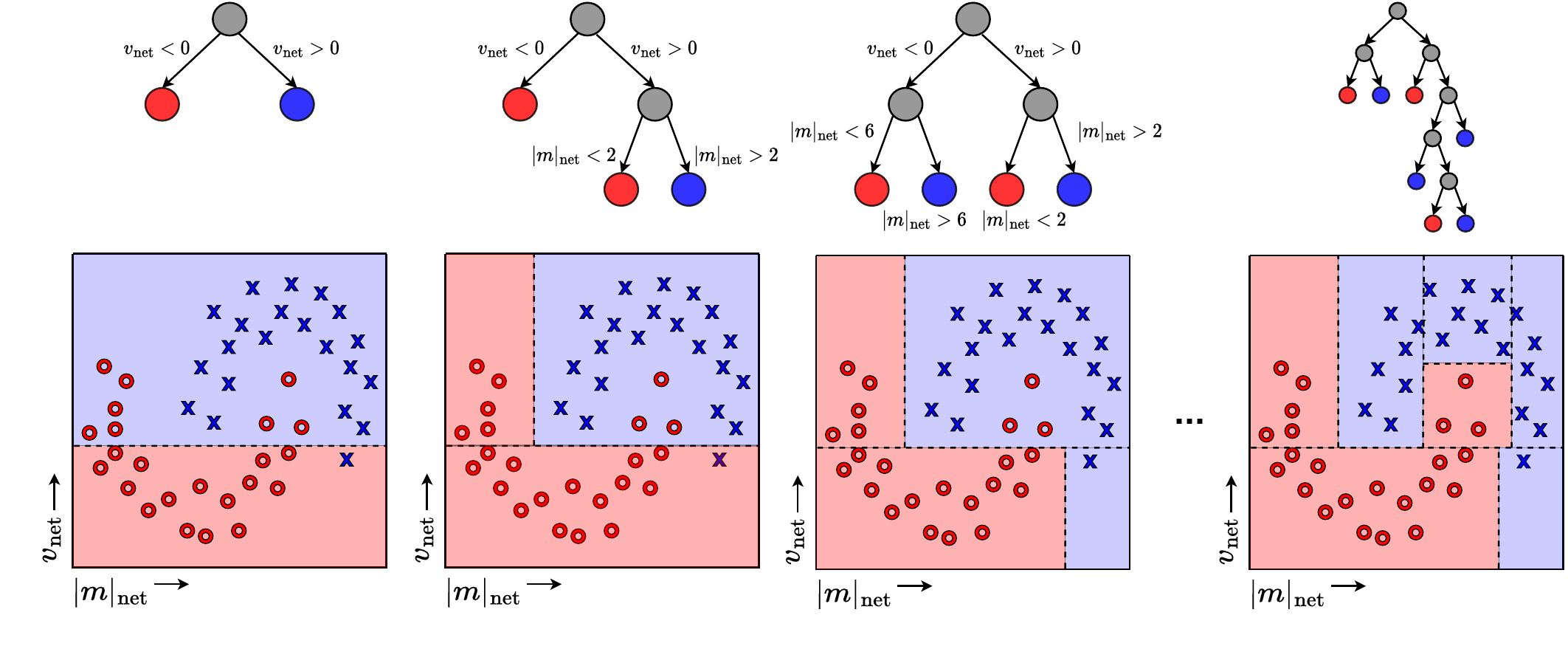}
  \caption{A schematic diagram of the iterative training process of a single decision tree on a toy dataset. At every iteration, the decision tree training algorithm considers adding a ``split'' in every variable of the input data set (in this case $|m|_\mathrm{net}$ or $v_\mathrm{net}$, but in general we consider all 11 observables) between all values of that variable. Across all variables and split locations, the decision tree chooses the split according to a particular criterion. For classification tasks, a typical criterion is maximizing the difference in Gini impurity before and after the split. For regression, minimizing the mean squared error is typical. After the split, the newly created leaf nodes are assigned new output values. The training algorithm iteratively splits nodes until reaching some stopping criteria. In this paper, we stop training when the number of data points represented by a given leaf node in the tree is less than 8. For simplicity, in this diagram, we show a binary classification task (categorizing data points as either red circles or blue crosses) but the algorithm generalizes in a straightforward way to regression. In particular, rather than having the output value of each leaf node be the majority class in a given region, the output value is the \textit{average} value of data points in a given region. 
  \label{fig:tree_training}}
\end{figure*}

A random forest is a collection of independently trained decision trees, which are commonly used in data mining to predict some target value from a set of input variables. In our case, the target value may be a quantity such as spin, that we would like to infer from the input variables which are our observables. Each tree is given a bootstrapped sample of the model images, which is then optimized to predict the target value.  A separate forest is constructed for each of the target values.  

We provide a schematic diagram of the training process for an individual tree in \autoref{fig:tree_training}, where training proceeds from left to right.  In this example, for simplicity, the target value is a boolean such as magnetic field polarity, represented by red circles and blue crosses.  We consider here only two input variables, $v_\mathrm{net}$ and $|m|_\mathrm{net}$, and the two magnetic field polarities occupy two complicated regions in this space.  The decision tree is trained by making successive cuts in the space spanned by the input variables until a termination condition is reached.  In computer science parlance, we start at the root node and branch the tree by making cuts until we reach a termination condition, after which each leaf node of the tree contains a prediction value for the target variable. Specifically, the training algorithm iteratively finds the next best cut among all leaf nodes and updates the tree by splitting the appropriate node. To do so, we consider all splits across all input variables, and add the split that maximises the tree's improvement in performance, measured in terms of mean squared predictive error (when performing regression) or Gini impurity \citep{breiman+1984} (when performing classification).  For classification tasks, the output value of a leaf node is a given class. For regression, the value is the average output value of training data points in the corresponding region of the data space.  In our example, the fully trained tree's predictions are perfect.

Each tree in the random forest is different because it has seen a different subset of the data.  When using the random forest to make predictions, the predictions of each tree are aggregated.  For classification, the aggregation method is majority vote, and for regression, the method is averaging. Collecting many individual learners to form a single, more robust algorithm in this way is called \textit{bootstrap aggregation}. The advantage of this method is that while each decision tree may overfit on its given training dataset, the collection of trees formed by bootstrap aggregation is robust to overfitting, given enough trees \citep{Breiman2001}. 

For our libraries, we partition each image time series into an 80 per cent - 20 per cent training-testing split, partitioned chronologically to avoid autocorrelation between training and testing datasets. On our training datasets, we fit random forests to predict spin, inclination (Sgr A* only), log-$R_\textrm{high}$, and $B$-field direction (for M87* only) using scikit-learn \citep{scikit-learn}. When predicting on spin, $R_\mathrm{high}$, and inclination, we used a regression forest and fit to minimise mean squared error. For predicting the magnetic field polarity, which is a boolean instead of a continuous quantity, we used a classification forest and used Gini impurity as our criterion.  We train forests with 400 trees each with a minimum of 8 data points to split a node. We selected each of these parameters following a grid-based parameter sweep but found that our results were robust to moderate deviations in the number of trees and minimum split.

\begin{figure*}
  \centering
  \includegraphics[width=\textwidth]{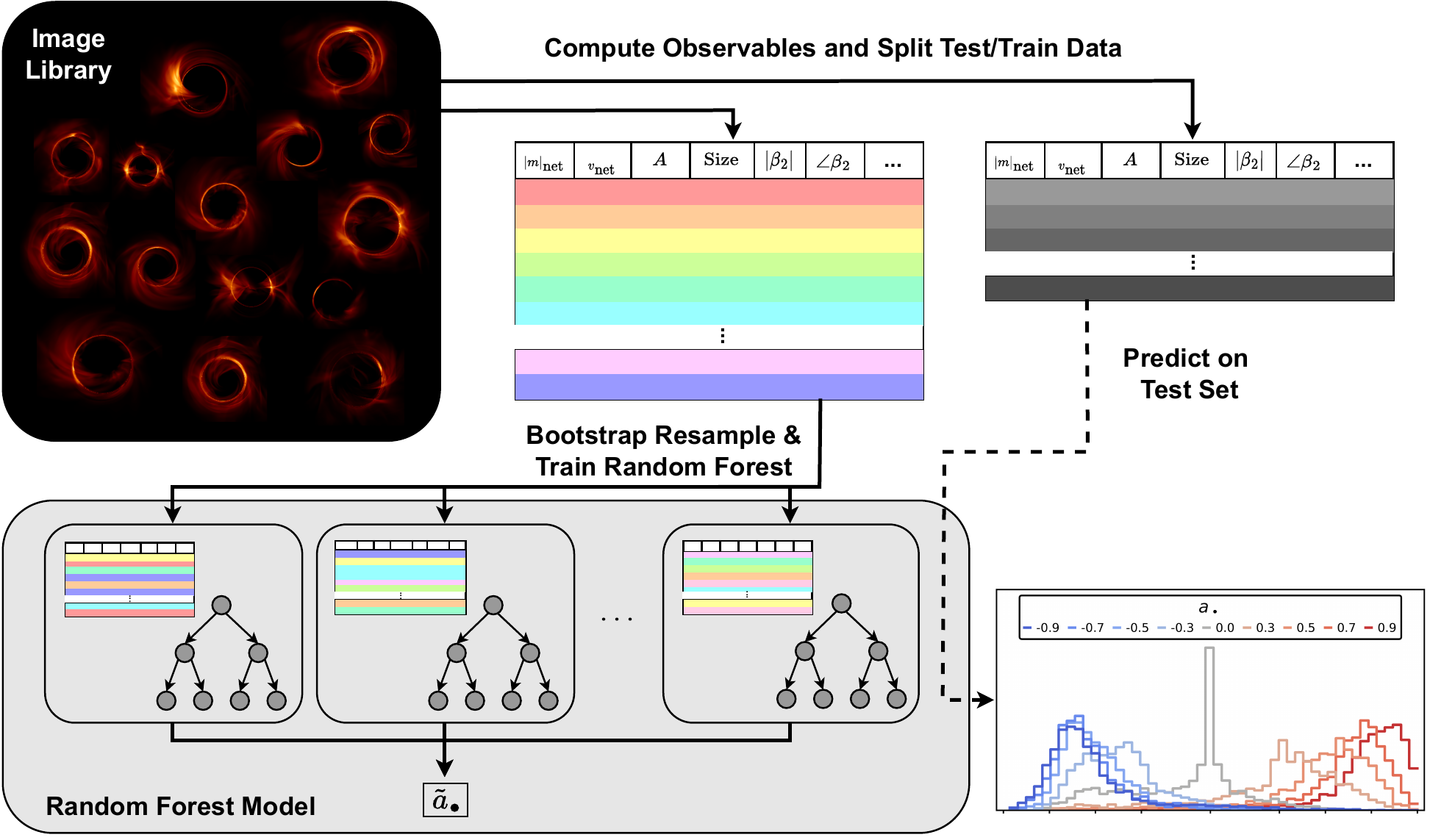}
  \caption{A schematic diagram of the methodology of \autoref{sec:library} and \autoref{sec:machine_learning}. Starting in the top left, we show that we have generated a large image library, as described in \autoref{sec:grrt}. Then, for each image, we compute a set of observables after blurring the image with a 20 $\mu$as Gaussian beam, as described in \autoref{sec:observables}. We split the images into training data and testing data for the machine learning model and use the former to train random forest models to predict quantities of interest (for instance $a_\bullet$), described in \autoref{sec:rfs}. Finally, we predict the same quantities of interest for the unseen test images and show the predicted distributions, per \autoref{sec:predictions}. 
  \label{fig:ml_overview}}
\end{figure*}

We present a schematic diagram of our methodology as a whole in \autoref{fig:ml_overview}.  First, an image library is created of M87* and Sgr A* models using the GRRT code {\sc ipole}.  From these images, we compute tables of observables that we believe are observable and robust.  We use the first 80 per cent of these values chronologically for training, and reserve the last 20 per cent for testing.  For the training set, we build random forests to predict spin, inclination, $R_\mathrm{high}$, and magnetic field polarity.  Finally, we apply these random forests to the unseen data to evaluate performance.

\subsection{Random Forest Model Performance}
\label{sec:predictions}

\begin{figure}
    \centering
    \includegraphics[width=0.45\textwidth]{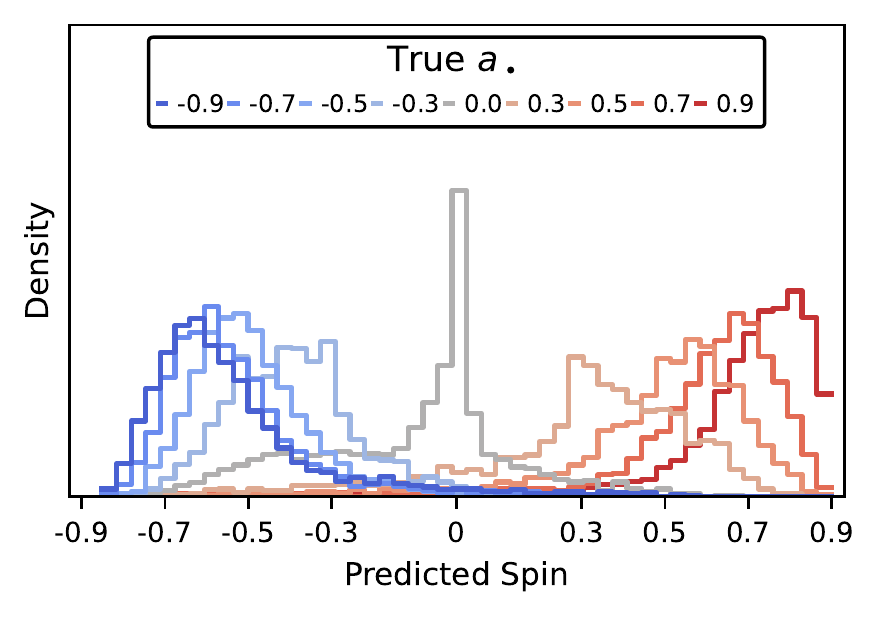}
    \caption{Predicted distributions of spin on unseen Sgr A* models by our random forest as a function of true spin (indicated by colour).  The model can successfully make meaningful predictions and almost always distinguishes progrades from retrogrades.  For high spins, the distributions are biased towards lower values, due in part to our method of averaging tree predictions.}
    \label{fig:postpred_spin}
\end{figure}

\begin{figure}
    \centering
    \includegraphics[width=0.45\textwidth]{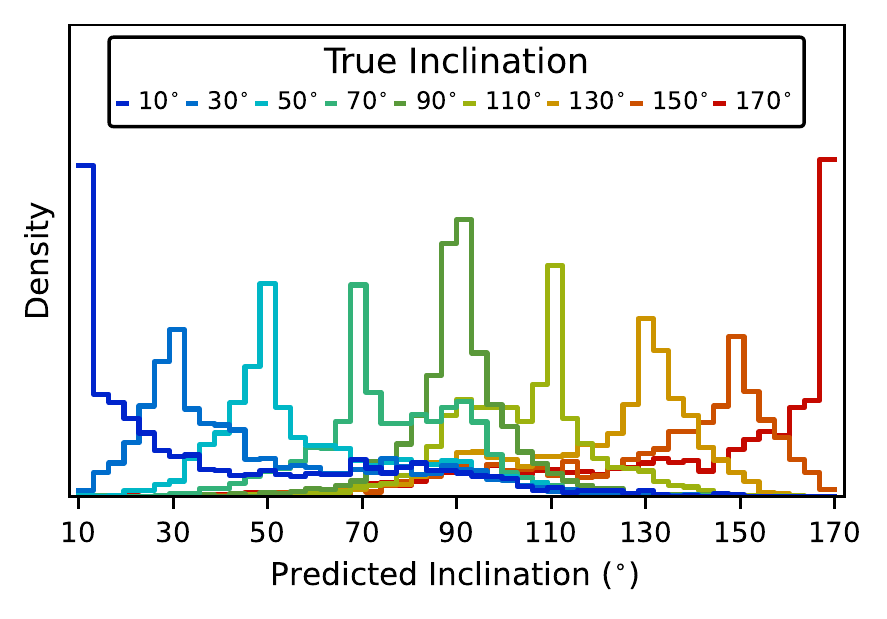}
    \caption{Distributions of predicted inclination for unseen data on the Sgr A* random forest model. Colour indicates true inclination.}
    \label{fig:postpred_inc}
\end{figure}

\begin{figure}
    \centering
    \includegraphics[width=0.45\textwidth]{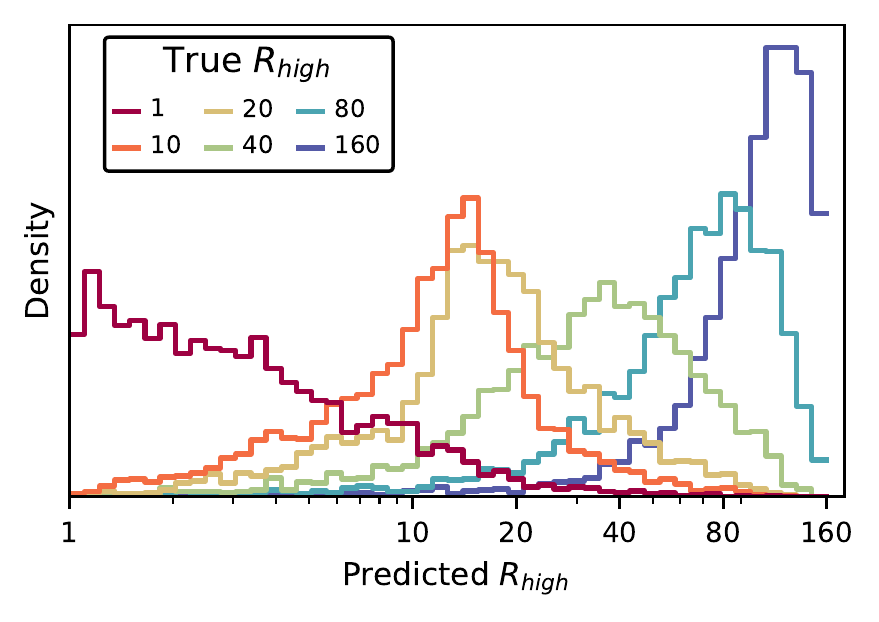}
    \caption{Distributions of predicted $R_\mathrm{high}$ for unseen data on the Sgr A* random forest model.  Colour indicates true $R_\mathrm{high}$ value, which is again well-predicted with biases towards more central values. We train and predict values of logarithmic $R_\mathrm{high}$, as reflected by the horizontal axis.}
    \label{fig:postpred_rhigh}
\end{figure}

\begin{figure}
    \centering
    \includegraphics[width=0.35\textwidth]{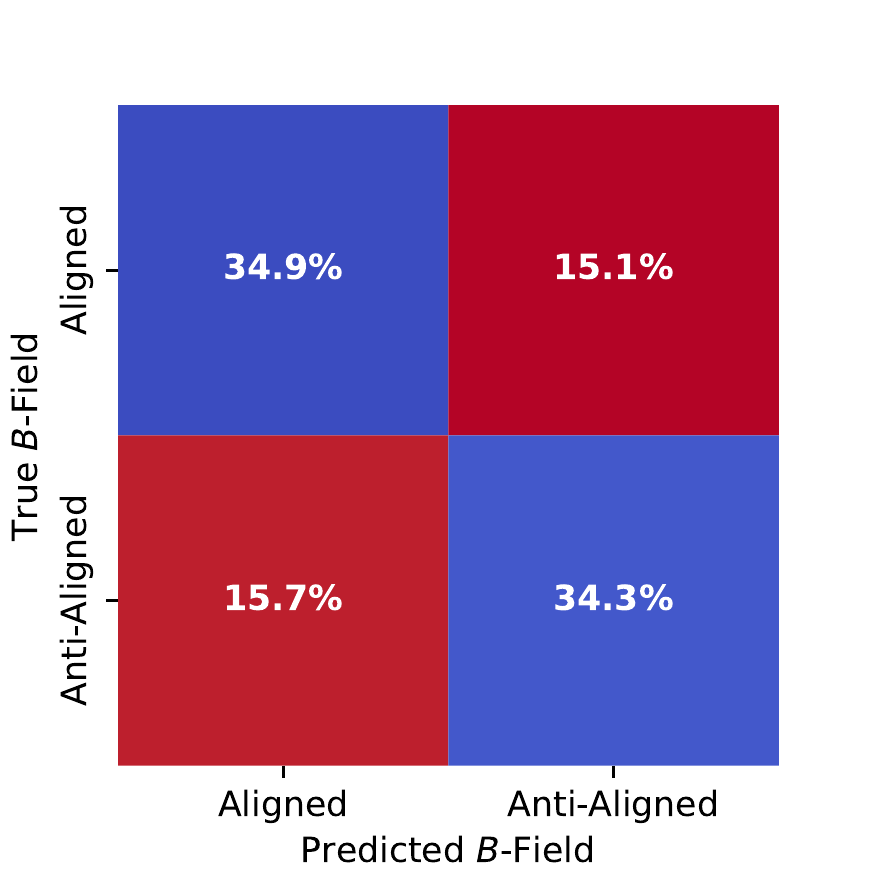}
    \caption{Categorical distributions of predicted $B$-field alignment for unseen data on the M87* random forest model. }
    \label{fig:postpred_bfield}
\end{figure}

After training a given model, we perform inference on our test library for the relevant predicted quantity to produce a predicted test distribution. We show predictive distributions for our Sgr A* library in spin, inclination, and $R_\textrm{high}$ in Figures \ref{fig:postpred_spin}, \ref{fig:postpred_inc}, and \ref{fig:postpred_rhigh}, respectively, and we show the predictive distribution for our M87* library in $B$-field direction in \autoref{fig:postpred_bfield}. As outlined in \autoref{sec:rfs}, each predictive distribution represents the output of a different random forest model for each inferred quantity, though each random forest sees the same set of computed observables for training. We note that in general, we expect that for extreme values in the predictive range (e.g. $a_\bullet = \pm 0.9$), we expect our models to systematically predict closer to the overall mean because the predicted values are an average of individual decision tree outputs which cannot exceed the extreme values in the predictive distribution. 

For spin, the predicted distributions for prograde, retrograde, and zero-spin models share very little overlap, indicating high model confidence in distinguishing each of these cases. However, there is substantial overlap among prograde and retrograde spins with different $|a_\bullet|$. Notably, the predicted distributions for $a_\bullet = \{-0.9, -0.7, -0.5\}$ are nearly indistinguishable, indicating that for Sgr A*, the model cannot discriminate well between high retrograde spin images, whereas it can discriminate more meaningfully between high prograde spins. We also find that our model performs slightly better for predicting spin at face-on inclinations than edge-on, likely because $\beta_2$ is stronger and more discriminating for different values of spin at face-on inclinations. 

Our random forest model is more successful at predicting inclination of our Sgr A* models.  In \autoref{fig:postpred_inc}, the predictive distributions for each inclination captured by our library have clearly distinguished peaks centreed at their true inclination values. However, each predictive distribution still contains a long tail which deviates substantially from the true value, likely due to time variability and turbulence in these models.

Our random forest model can also meaningfully infer $R_\mathrm{high}$, shown in \autoref{fig:postpred_rhigh}. As mentioned in \autoref{sec:rfs}, we train and predict on log-$R_\mathrm{high}$, reflected in \autoref{fig:postpred_rhigh}. The predictive distributions contain substantial overlap with adjacent values of $R_\mathrm{high}$ but large differences in $R_\mathrm{high}$ are well distinguished (for instance, $R_\mathrm{high} = 10$ and $R_\mathrm{high} = 80$). We note that the wide distribution of $R_\mathrm{high} = 1$ is likely because our library does not contain intermediate values between $R_\mathrm{high} = 1$ and $R_\mathrm{high} = 10$ which is a larger gap in log-space than the rest of the values of $R_\mathrm{high}$ captured by our library. 

Lastly, in \autoref{fig:postpred_bfield}, we plot the binary predictive distributions of the $B$-field direction. Unlike the other models, which perform regression, our model for predicting the $B$-field alignment performs binary classification. Our model is weakly able to discriminate $B$-field direction, predicting the correct alignment for about $70\%$ of test images. 

\subsection{Interpreting Model Predictions}
\label{sec:shap_values}

We can use machine learning interpretability tools to provide explanations for our individual predictions. SHapely Additive exPlanations (SHAP) is a recently proposed interpretability tool which has quickly gained popularity \citep{lundberg+2017}. SHAP borrows ideas from coalition game theory to unify several existing frameworks for local model interpretability while improving theoretical guarantees on the explanations.

Briefly, SHAP unpacks individual model predictions by attributing a \textit{SHAP value} to each feature (observable, in our case) for every model prediction. A feature's SHAP value represents its contribution towards the overall model prediction.  The SHAP values across all features for a given prediction sum to the difference between (i) the mean prediction over the entire dataset (e.g. $a_\bullet = 0$ or $i = 90^\circ$) and (ii) the model prediction on that individual image. 

Consider a model $f(x)$ that makes a prediction as a function of data $x$.  Let $R$ be an ordering of features, and let $\mathcal{R}$ be the set of all possible orderings.  The SHAP value of feature $i$ on model $f$ at data point $x$ is given as follows \cite{lundberg+2017}: 
\begin{equation}
    \mathrm{SHAP}_i(f, x) = \frac{1}{|\mathcal{R}|} \sum_{R \in \mathcal{R}}  \left[ f_x(P_i^R \cup i) - f_x(P_i^R) \right]
\end{equation}
where $P_i^R$ is the set of features in ordering $R$ that come before (and exclude) feature $i$, $P_i^R \cup i$ is the set of all features that come before and include feature $i$, and $|\mathcal{R}| = M!$, where $M$ is the number of input features of the model. $f_x(S)$ for some set of features $S$ is given by $f_x(S) = E\left[f(x) | x_S \right]$, where $f(x)$ is the model prediction on data point $x$ and $x_S$ is the subset of $x$ with only features in $S$; thus $f_x(S)$ is the average model prediction on data point $x$ with only the features in $S$. For tree-based models, $f_x(S)$ is calculated in low-order polynomial time by the algorithm given in \cite{lundberg+2020}. 

Heuristically, the SHAP value for a given feature and prediction is how much the individual feature value ``pushes'' the overall model prediction on a given image relative to predictions without that feature. For example, consider our model trained on our Sgr A* library to predict spin. For every image in the test library, we can compute a SHAP value for each observable. If a given image has $\angle\beta_2 = -\pi/2$ and we calculate that the $\angle\beta_2$ SHAP value is 0.4, then compared to model predictions on subsets of input features without $\angle\beta_2$, adding $\angle\beta_2 = -\pi/2$ to the set of input features increases our model's predicted value of spin by an average of 0.4 for that image.   The larger the magnitude of the SHAP value is for a given feature and prediction, the more important that feature is towards the final model output. In \autoref{sec:feat_imps}, we discuss using the mean absolute SHAP value of a given feature across the test dataset as a feature importance measure in more detail.  For more details, we refer readers to \cite{lundberg+2017} and \cite{molnar2022}. Lastly, we note that SHAP values are in general exponentially costly to calculate for any given machine learning algorithm. However, for tree-based algorithms, they can be computed in polynomial time using the TreeSHAP algorithm \citep{lundberg+2020}, which exploits the branching structure of the trees to cache and avoid redundant computations.

\begin{figure}
    \centering
    \includegraphics[width=0.45\textwidth]{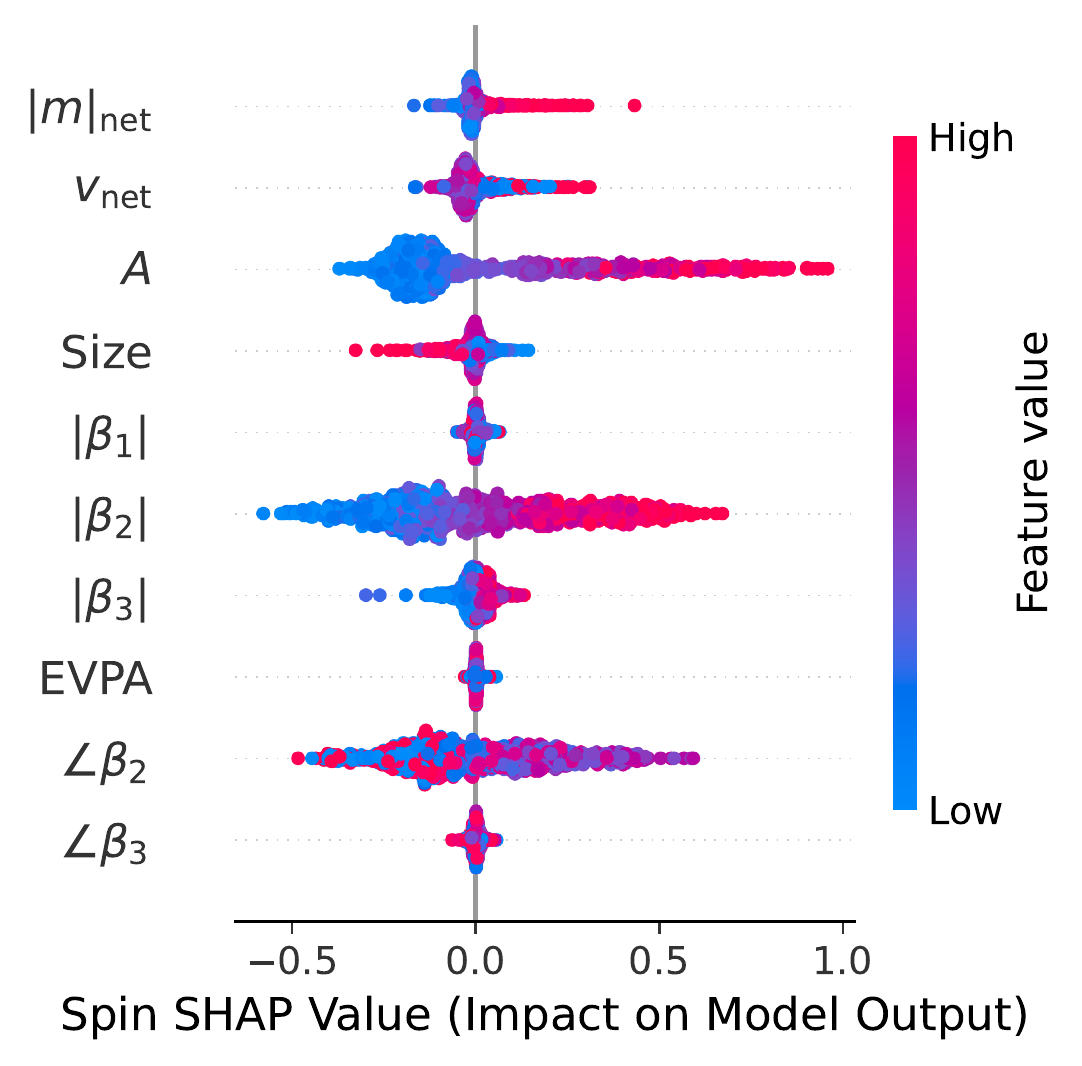}
    \caption{
    For each image feature, we visualise the distribution of its SHAP values for predicting spin.  Each point represents one image in the testing set of the Sgr A* library.  Since each feature has different numerical values, colours encode the relative value of each feature compared to the full range that it spans.  For example, for $\angle \beta_2$, the bluest points correspond to $-\pi$ and the reddest points correspond to $\pi$.  As discussed in the text, the larger the SHAP value, the larger impact of this value on predicting the spin.  The vertical width of each distribution corresponds to the frequency of images with that particular SHAP value.
    \label{fig:sgrA_spin_shap_summary}}
\end{figure}

We show distributions of test library SHAP values for each feature for our Sgr A* spin model in \autoref{fig:sgrA_spin_shap_summary}. We remind readers that, as described in \autoref{sec:observables}, since the orientation of the spin axis of Sgr A* projected on the sky is presently unknown, we anchor $\angle\beta_{0, 2, 3}$ to $\angle\beta_1$, and thus $\angle\beta_1$ is omitted from \autoref{fig:sgrA_spin_shap_summary}. Note that unlike in  \autoref{fig:histograms_sgrA_spin}, where we fix a particular value of viewing inclination and $R_\mathrm{high}$, \autoref{fig:sgrA_spin_shap_summary} includes SHAP values across the full parameter space of the library. Thus, these trends are not specific to particular inclinations or $R_\mathrm{high}$ values. Each feature has different numerical values, so they are colour-coded based on their position within the full range of values spanned by that feature in the test library.  Some features have SHAP values clustered around 0, and therefore do not have much impact on a spin prediction, such as $|\beta_1|$, EVPA, and $\angle \beta_3$.  On the other hand, $\beta_2$ and $A$ stand out as important due to their comparably large SHAP values.  As shown in \autoref{fig:histograms_sgrA_spin} and \citet{Medeiros+2022}, we generally expect that large values of asymmetry correspond to high prograde spins due to Doppler beaming.  Meanwhile, $\beta_2$ encodes both the magnetic field's geometry and relative order in its phase and amplitude respectively.  

\begin{figure}
    \centering
    \includegraphics[width=0.45\textwidth]{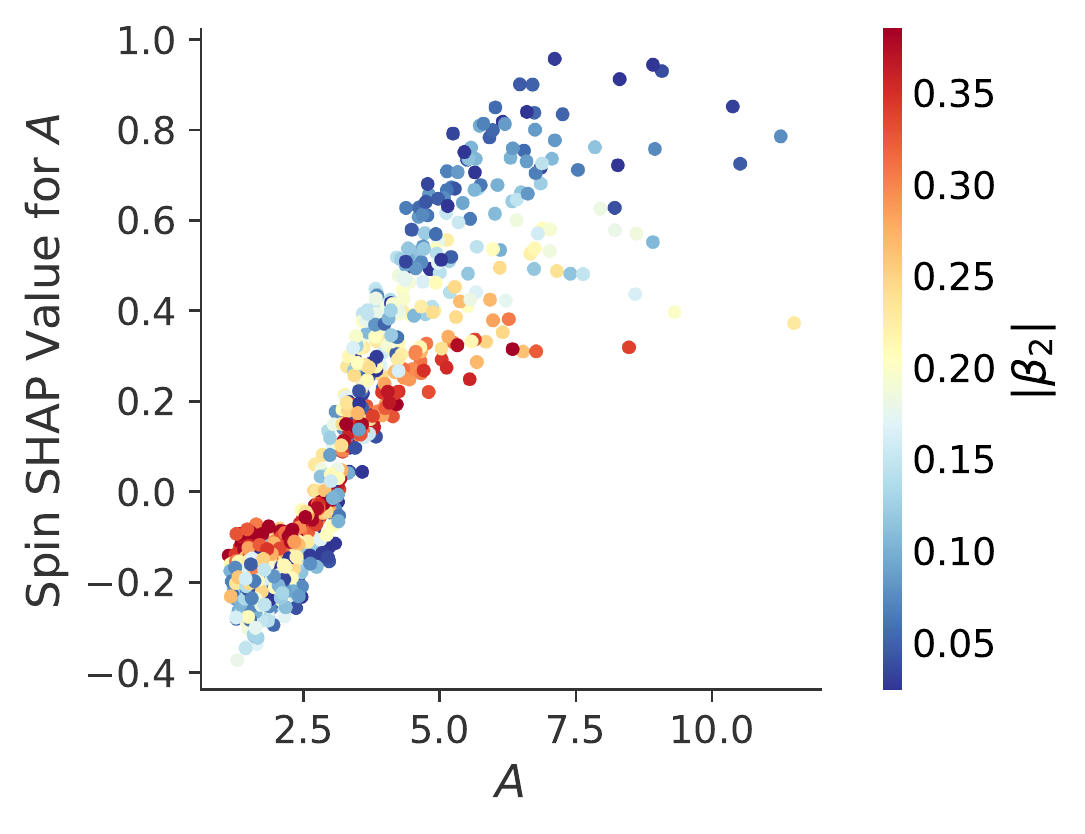}
    \caption{Distribution of SHAP values for predicting spin, showing the interaction between asymmetry and $|\beta_2|$. Large values of asymmetry push the prediction towards higher values of spin, due to Doppler beaming.  Then, for a given value of asymmetry, larger values of $|\beta_2|$ imply larger spin values for more symmetric systems, or lower spin values for more symmetric systems.
    \label{fig:sgrA_spin_shap_asymmetry}}
\end{figure}

We dig deeper into the interaction between $A$ and $|\beta_2|$ in \autoref{fig:sgrA_spin_shap_asymmetry}, where we plot the distribution of spin SHAP values as a function of $A$, colour-coded by $|\beta_2|$. We observe that asymmetry SHAP values appear to increase monotonically with asymmetry, with more common low asymmetry values having moderately negative SHAP values amd less common large asymmetry values having high positive SHAP values. This indicates that while the bulk of images have moderate asymmetry values that the model weakly assoicates with more negative spins, very high asymmetry values are associated with substantially more positive spin predictions.  However, high-asymmetry values can occur across a large range of prograde spins.  We see that $|\beta_2|$ seems to mediate the SHAP value for a given asymmetry value.  For small asymmetry values, low values of $|\beta_2|$ push for more negative predictions of spin, but this relationship inverts for large asymmetry values, where  lower values of $|\beta_2|$ result in more positive predictions of spin. 

\begin{figure}
    \centering
    \includegraphics[width=0.45\textwidth]{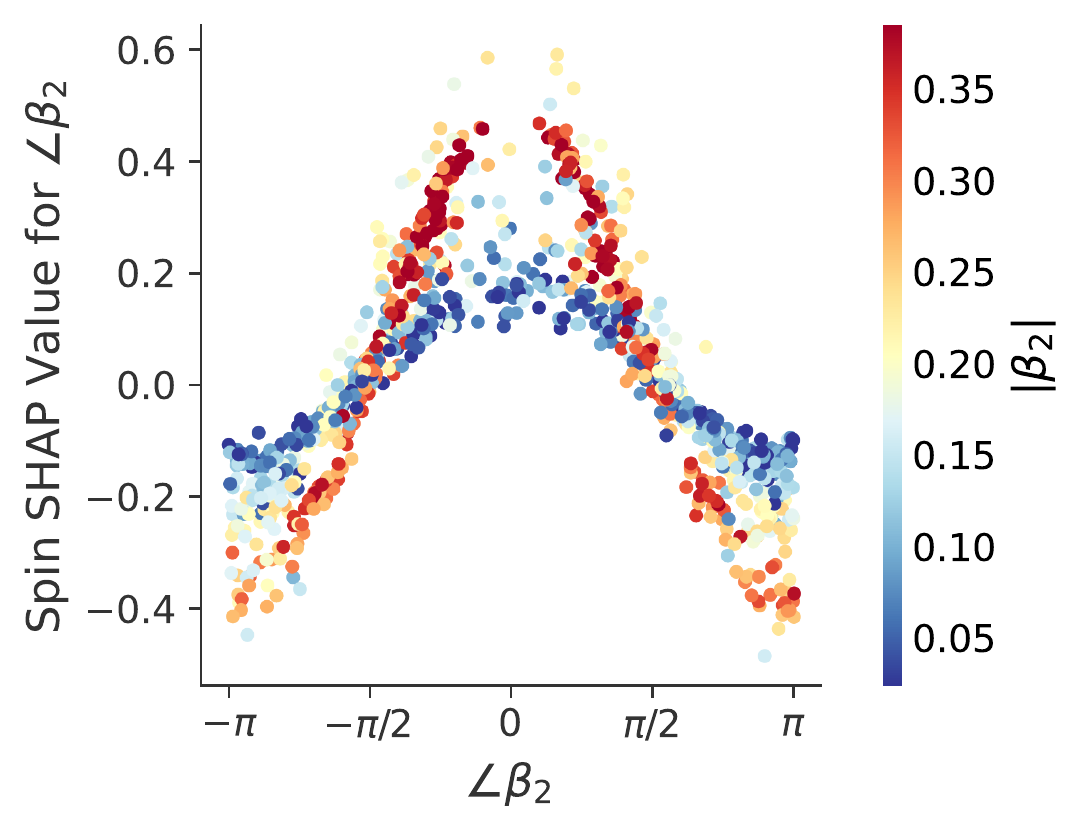}
    \caption{Distribution of SHAP values for predicting spin, showing the interaction between $\angle\beta_2$ and $|\beta_2|$.  The model learns to associate images with $\angle\beta_2 \approx 0$ (more radial EVPA patterns, and therefore more toroidal magnetic fields) with prograde spins and $\angle\beta_2 \approx \pm \pi$ with retrograde spins (more toroidal EVPA patterns, and therefore more radial magnetic fields).  $|\beta_2|$ carries similar information, since it declines for messy retrograde systems. 
    \label{fig:sgrA_spin_shap_beta2arg}}
\end{figure}

In \autoref{fig:sgrA_spin_shap_beta2arg}, we now examine the interaction between $\angle \beta_2$ and $|\beta_2|$.  We see a wide distribution of SHAP values, with more extreme phases (i.e. closer to $-\pi$ and $\pi$) yielding negative SHAP values and more moderate phases (i.e. closer to $\angle\beta_2 = 0$) giving more positive SHAP values. Referring back towards \autoref{fig:histograms_sgrA_spin}, this model inference stems from the phenomenon that high prograde spin images have narrow $0$-centred distributions in $\angle\beta_2$ whereas high retrograde spin models have distributions centred more closely around $\angle\beta_2 = \pm \pi$.  

\begin{figure}
    \centering
    \includegraphics[width=0.45\textwidth]{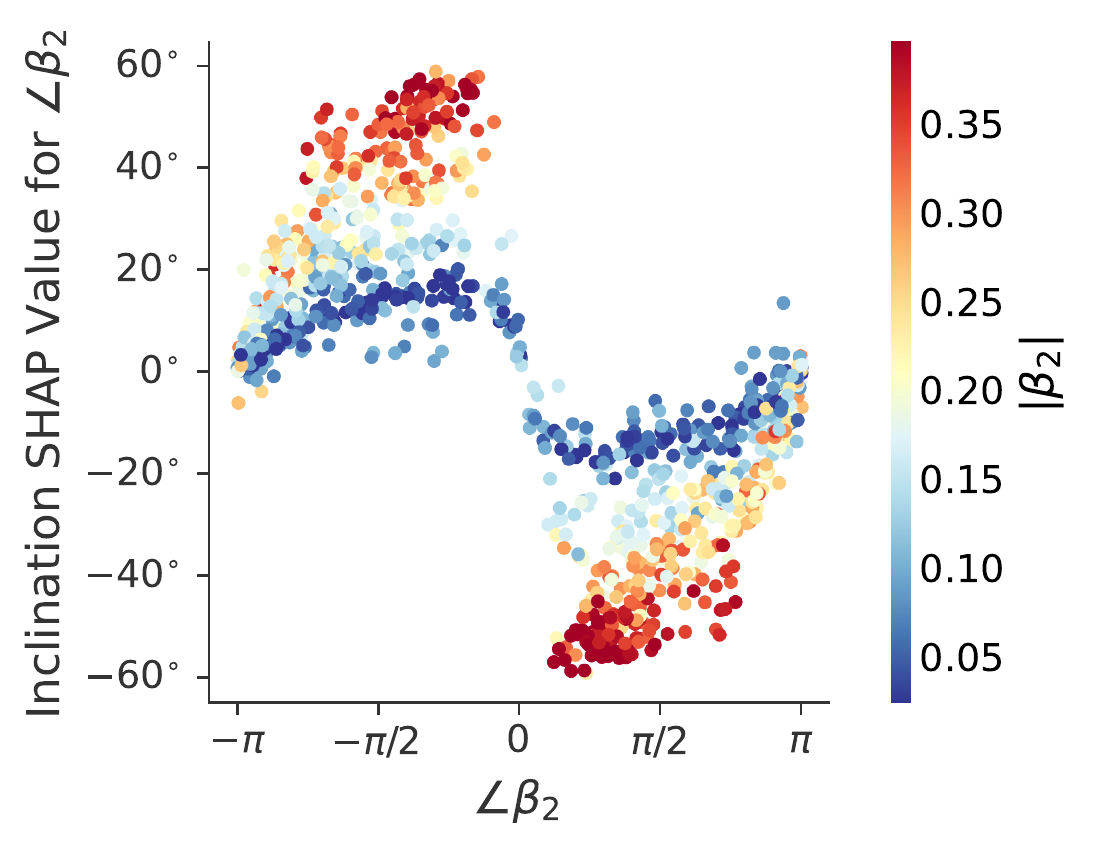}
    \caption{Distribution of SHAP values for predicting inclination, showing the interaction between $\angle \beta_2$ and $|\beta_2|$.  First, the model can use the sign of $\angle \beta_2$ to distinguish whether or not the inclination is greater than or less than the mean value of $90^\circ$.  This is because by construction the sign of $\angle \beta_2$ distinguishes clockwise from counterclockwise linear polarization ticks.  Then, larger values of $|\beta_2|$ imply more face-on viewing angles.  This is because face-on viewing angles result in more rotationally symmetric images.
    \label{fig:sgrA_inc_shap_beta2_arg}}
\end{figure}

Finally, in \autoref{fig:sgrA_inc_shap_beta2_arg}, we examine the relationship between $\angle\beta_2$ and $|\beta_2|$ for predicting inclination.  We observe a distinct bifurcation based on the sign of $\angle\beta_2$. For $\angle\beta_2 > 0$, the corresponding SHAP values are negative, meaning the model is pushed towards predicting inclinations closer to $0^\circ$. Conversely, for $\angle\beta_2 < 0$, the SHAP values are positive, and the model prefers predicting inclinations near $180^\circ$.  This is because the sign of $\angle \beta_2$ directly encodes the handedness of the spiral EVPA structure, which flips if you flip the viewing angle.  Further, the gradient of $|\beta_2|$ demonstrates that larger values of $|\beta_2|$ lead to more face-on inclination predictions. This statistical relationship learned by our model is consistent with the trend observed in the distribution of quantities as a function of inclination (see \autoref{fig:histograms_sgrA_incs}) and for which we provide physical motivation in \autoref{sec:obs_distributions}. 

\subsection{Feature Importances}
\label{sec:feat_imps}

\begin{figure}
    \centering
    \includegraphics[width=0.45\textwidth]{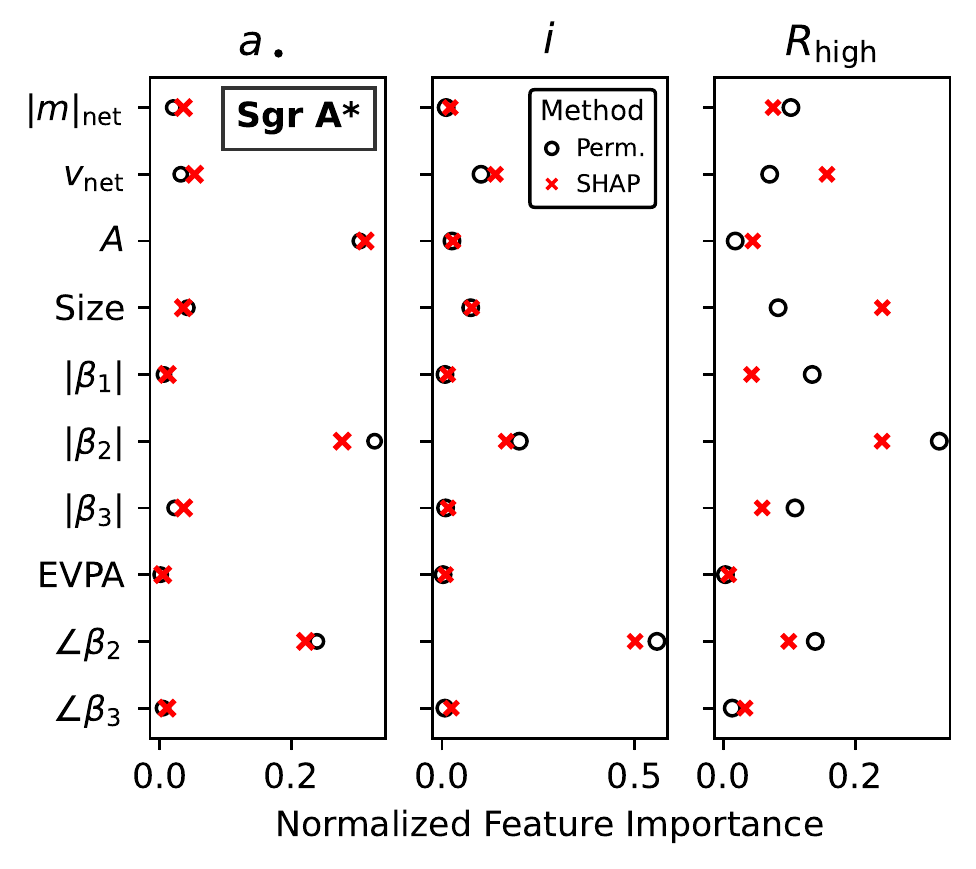}
    \caption{Permutation and SHAP feature importances among our Sgr A* models for inferring $a_\bullet$, $i$, and $R_\mathrm{high}$, where larger values correspond to greater importance. Here, feature importance values are normalised such that the sum of all feature importances for a given model and method sum to unity. 
    \label{fig:sgrA_feature_importances}}
\end{figure}

\begin{figure}
    \centering
    \includegraphics[width=0.45\textwidth]{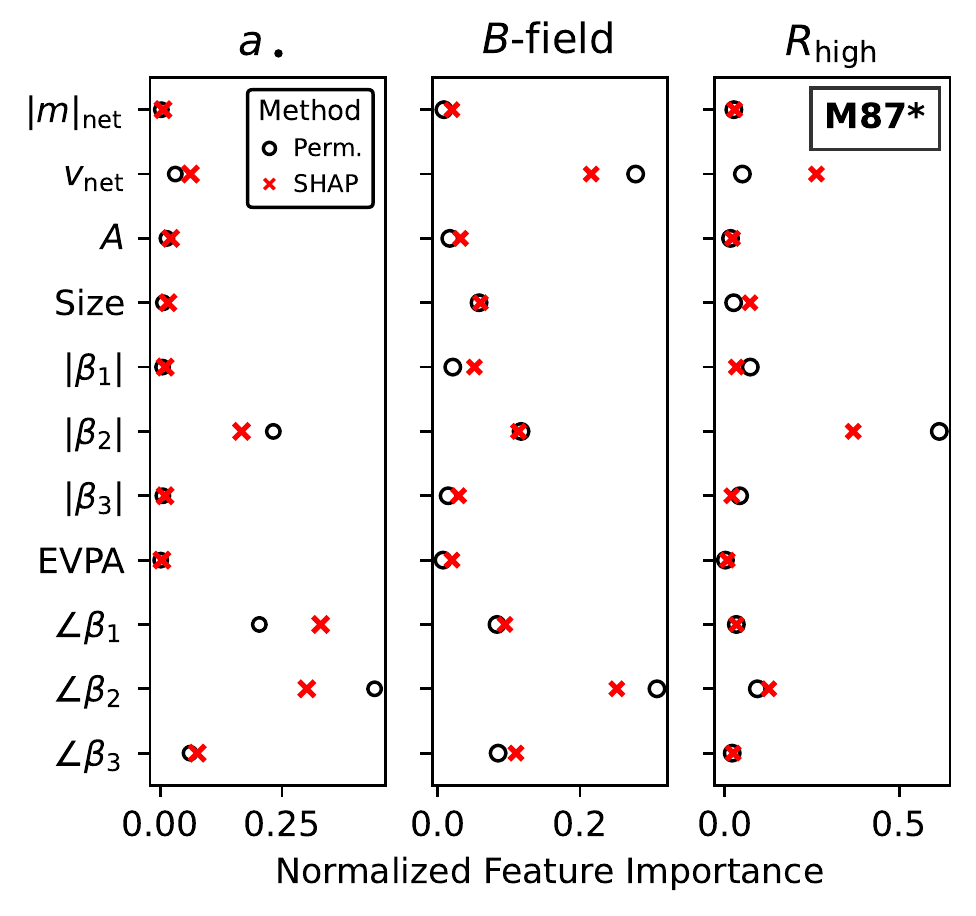}
    \caption{Permutation and SHAP feature importances among our M87* models for inferring $a_\bullet$, the magnetic field alignment, and $R_\mathrm{high}$, where larger values correspond to greater importance. Here, feature importance values are normalised such that the sum of all feature importances for a given model and method sum to unity. 
    \label{fig:m87_feature_importances}}
\end{figure}

In addition to explanations of individual predictions based on a given image's computed observable values, we can also examine the distribution of predictions and assess the overall importance of each image observable across testing libraries for each model. In particular, we consider SHAP feature importances and permutation feature importances. 

To compute the importance of a feature using SHAP values, we compute SHAP values for that feature on every image in the testing distribution (e.g. \autoref{fig:sgrA_spin_shap_summary}). Then, the feature importance is the mean absolute SHAP value for that feature across the test library \citep{lundberg+2020}. Intuitively, this is a sensible metric as large SHAP values correspond to a given feature having a large influence on the model prediction for an individual image. 

As a simpler alternative, we also calculate a permutation feature importance \citep{Breiman2001} for each library and image observable. To do so, we first compute a reference pseudo-$R^2$ score over the set of testing predictions, defined as: 
\begin{equation}
    R^2 = 1 - \frac{\sum_j (y_j - \hat{y}_j)^2}{\sum_j (y_j - \overline{y})^2}
\end{equation}

\noindent where $j$ indexes over images, $y_j$ is the true value (e.g. the true spin or true inclination) for image $j$, $\hat{y}_j$ is the model prediction on image $j$, and $\overline{y}$ is the mean value of $y_j$ over all images. To compute the permutation feature importance for observable quantity $x$, we randomly shuffle the order of the computed $x$ in the test library while holding the remaining observables fixed.  Then, we compute the new $R^2$ value for our shuffled dataset and calculate the feature importance as the decrease from our reference $R^2$.  Essentially, this procedure tests how much the model's predictions worsen if one of the features is removed.

In \autoref{fig:sgrA_feature_importances}, we show normalised feature importances (such that the sum of importances sum to 1 for SHAP and permutation methods) for each Sgr A* model in predicting spin, inclination, and $R_\textrm{high}$. We remind readers that, as in \autoref{fig:sgrA_spin_shap_summary} and described in \autoref{sec:observables}, since the orientation of the spin axis of Sgr A* projected onto the sky is presently unknown, we anchor $\angle\beta_{0, 2, 3}$ to $\angle\beta_1$, and thus $\angle\beta_1$ is omitted from \autoref{fig:sgrA_feature_importances}. For predicting spin, $|\beta_2|$ and $\angle \beta_2$ have the greatest feature importances, followed by asymmetry and size. The remaining features have relatively low importances. The reason for these feature importance trends is evident in \autoref{fig:histograms_sgrA_spin}, where we observe that both $|\beta_2|$ and $\angle \beta_2$ trend with $a_\bullet$ with fairly tight distributions with minimal overlap for different values of $a_\bullet$. Similarly, the distribution of $A$ gets wider at higher prograde spins, though with more overlap at differing values of $a_\bullet$. 

For inclination, \autoref{fig:sgrA_feature_importances} shows that $\angle \beta_2$ is highly important,  $|\beta_2|$ and $v_\mathrm{net}$ are moderately important, and all other observables are relatively unimportant. Looking towards Figure \ref{fig:histograms_sgrA_incs}, we see that sign of $\angle \beta_2$ cleanly indicates whether $i > 90^\circ$ or $i < 90^\circ$. This is the largest segregation in terms of absolute error so both feature importance methods rank $\angle\beta_2$ as the most important observable. We also observe that $v_\mathrm{net}$ plays a similar role but does not discriminate as cleanly. Once $i > 90^\circ$ or $i < 90^\circ$ is known, $|\beta_2|$ can be used to determine $|i - 90^\circ|$, which follows from the distributions of observables shown in \autoref{fig:histograms_sgrA_incs} and our examination of SHAP values in \autoref{sec:shap_values} (particularly \autoref{fig:sgrA_inc_shap_beta2_arg}). 

Looking towards $R_\textrm{high}$ we find that $|\beta_2|$ and $\angle\beta_2$ remain among the most important features. \autoref{fig:histograms_sgrA_rhigh} suggests that $|\beta_2|$ in particular trends strongly with $R_\mathrm{high}$.  As also explored in \citet{EHTC+2021b}, models with larger $R_\mathrm{high}$ have larger Faraday rotation depths, leading to more scrambling, and therefore smaller $|\beta_2|$.  However, we also find more disagreement between our two feature importance methods and that generally more features are important. Since the ``importance'' of a feature is not well-defined, we do not necessarily expect agreement between differen feature importance methods. In this case, the disagreement may suggest that  inferring $R_\mathrm{high}$ is a more difficult and messy problem, and that $R_\mathrm{high}$ changes images in many different ways. The SHAP distributions of $R_\mathrm{high}$ (not shown) suggest our model learns to infer complex higher-dimensional relationships among input features, some of which are likely used to indirectly infer the spin and inclination of a particular image. 

In \autoref{fig:m87_feature_importances}, we show permutation and SHAP feature importances for each of our M87* models predicting spin, $R_\mathrm{high}$, and $B$-field direction.  For $a_\bullet$, we observe similar trends to Sgr A* with the exception that the asymmetry $A$ is no longer important. This is because our M87* model is fixed at observing inclinations of $163^\circ$ and $17^\circ$ and asymmetry discriminates most strongly for inclinations closer to edge-on. For M87*, the distributions of asymmetry as a function of spin overlap nearly entirely, unlike that for Sgr A* even at moderately edge-on inclinations, such as that shown in \autoref{fig:histograms_sgrA_spin}. We also discover that $\angle\beta_1$ plays an important role in discriminating spin for M87*. This is because we observe a shift in $\angle\beta_1$ as a function of inclination (exemplified by the same in Sgr A*, shown in \autoref{fig:histograms_sgrA_incs}), which for our M87* library indirectly distinguishes progrades from retrogrades. 

For $R_\mathrm{high}$, we observe fewer highly important features compared to Sgr A* but similar general trends. In particular, $|\beta_2|$ remains the most important feature. However, for M87*, again because we fix the inclination, we find that $|\beta_2|$ and $v_\mathrm{net}$ are alone sufficient to capture a strong trend in $R_\mathrm{high}$ without needing to implicitly infer intermediate quantities such as viewing inclination as is the case for Sgr A*.

Finally, for the magnetic field alignment, we find that $\angle\beta_2$ and $v_\mathrm{net}$ are the most important quantities. This is unsurprising, as these two quantities track most strongly with magnetic field alignment, as shown in \autoref{fig:histograms_m87_bfield}. The model finds $\angle\beta_2$ more important that $v_\mathrm{net}$, likely because the discrimination between the distributions is more distinct for $\angle\beta_2$. 

\subsection{Repeated Polarimetric Observations}
\label{sec:repeat_obs}

\begin{figure*}
    \centering
    \includegraphics[width=\textwidth]{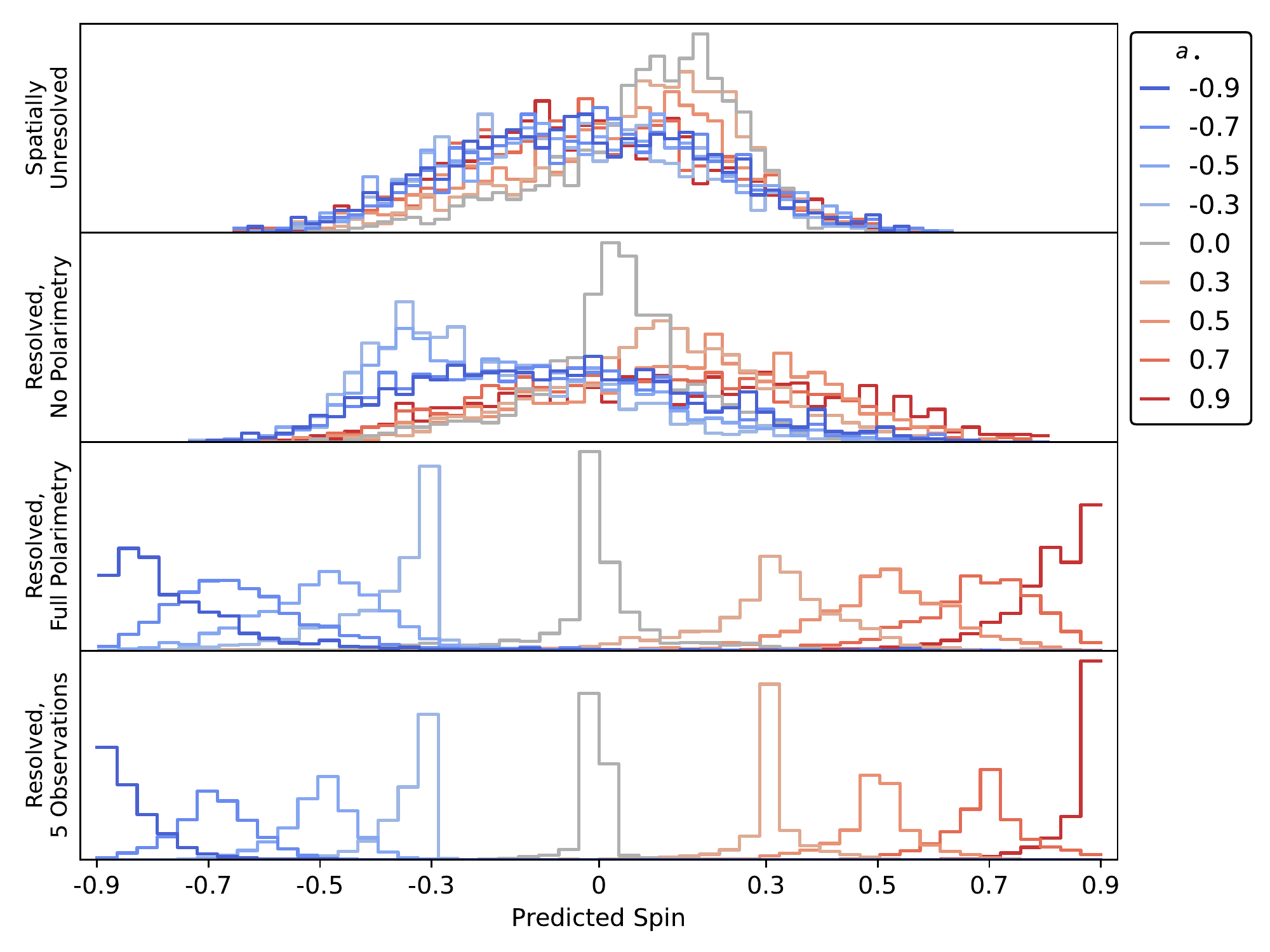}
    \caption{Random forest predicted distributions of spin for increasing amounts of information available from our M87* models, as described in more detail in \autoref{sec:repeat_obs}.  Spatially resolved polarimetric information (panels 3 and 4), in particular $\beta_2$, is essential for inferring spin, since this encodes the magnetic field geometry.  All observables fluctuate around their mean values due to turbulence in the accretion flow.  Predicted distributions grow narrower if multiple independent epochs are observed (panel 4), motivating continued monitoring of EHT sources.}
     \label{fig:m87_addinfo}
\end{figure*}

So far, all of our model predictions have involved only a single snapshot of information.  The EHT has already observed its targets for multiple (unpublished) epochs and will continue to do so.  We train models on varying amounts of sampled information to simulate the effect of varying EHT resolving power and polarimetric capabilities. Specifically, we simulate the following scenarios with our M87* library:
\begin{itemize}
    \item Spatially unresolved: Only $|m|_\mathrm{net}$ and $v_\mathrm{net}$ included. 
    \item Resolved, no polarimetry: $|m|_\mathrm{net}$, $v_\mathrm{net}$, asymmetry, and second image moment. 
    \item Resoled, full polarimetry: All image features, viz., $|m|_\mathrm{net}$, $v_\mathrm{net}$, asymmetry, second image moment, $|\beta_j|$ and $\angle\beta_j$, $j \in \{0, 1, 2, 3\}$.
    \item Resolved, full polarimetry, 5 observations: All image features as above but with 5 random snapshots drawn from the same library. Independent samples are drawn with replacement and without regard to order.
\end{itemize}

Predictions on the corresponding test libraries are shown in \autoref{fig:m87_addinfo}. With spatially unresolved information alone, the model can produce no meaningful predictions on spin. With resolved images, the model begins to weakly separate spin-zero images from prograde images, and can more strongly separate out prograde and retrograde images. By introducing spatially resolved polarimetry, suddenly the model can distinguish different individual spins but with substantial overlap in the predicted test library distributions. Note that these predictions are much better than for Sgr A* (\autoref{fig:postpred_spin}) due to the known inclination of M87*. Finally, with 5 observations, we see substantially more distinct separation between spins, motivating repeated EHT observations of M87* and Sgr A*. 

\subsection{Interpreting Observations of M87*}

\begin{figure}
    \centering
    \includegraphics[width=0.49\textwidth]{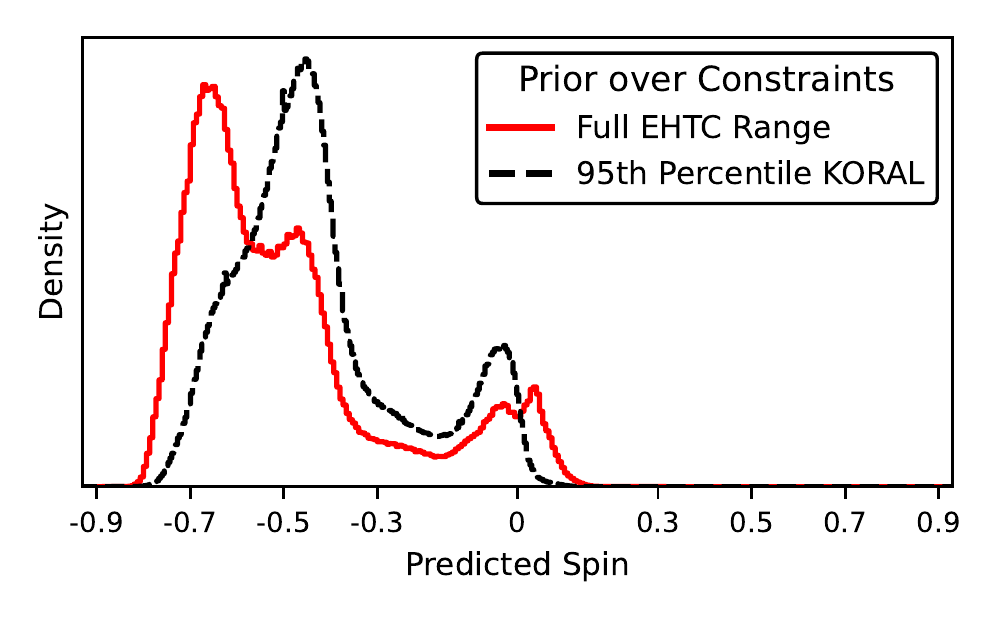}
    \caption{Posterior distributions of spin from independent uniform priors on observed M87* constraints \citep{EHTC+2021a} (red) and further constrained to lie within the 95th percentile of our M87* image library (dashed black).  Our model prefers a highly spinning retrograde spin for M87* for both priors.  Note that these values are likely biased low due to our averaging method.  The small peak at $a_\bullet=0$ could likely be eliminated by applying a jet power constraint. 
    \label{fig:m87_pred_spin}}
\end{figure}

\begin{figure}
    \centering
    \includegraphics[width=0.49\textwidth]{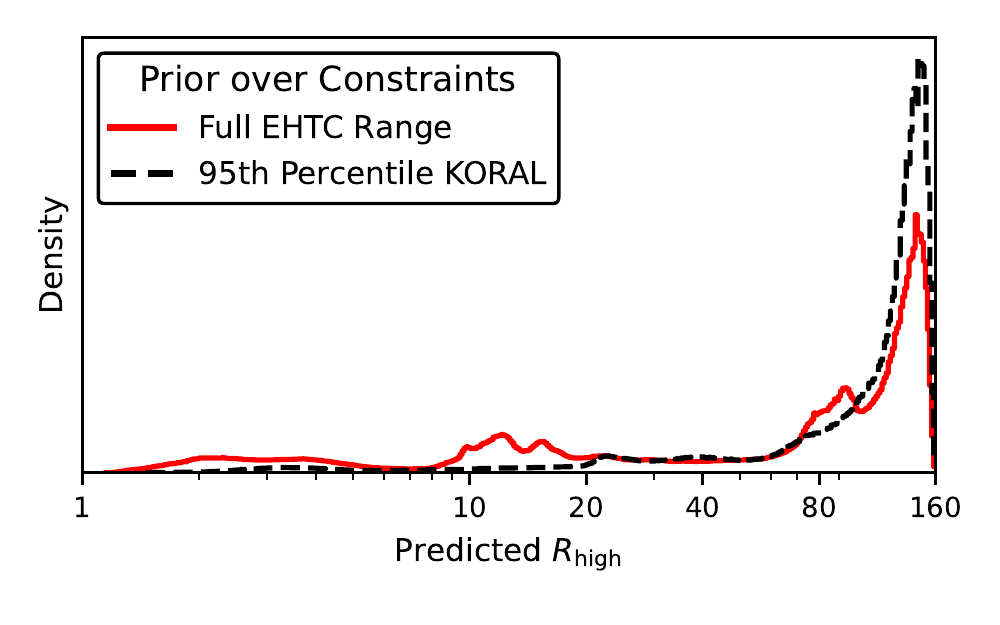}
    \caption{Posterior distributions of $R_\textrm{high}$ from independent uniform priors on observed M87* constraints \citep{EHTC+2021a} (red) and further constrained to lie within the 95th percentile of our M87* image library (dashed black).  Our methodology prefers large values (much colder electrons than ions), particularly with our 95th percentile prior, which helps depolarize models via Faraday rotation. 
    \label{fig:m87_pred_rhigh}}
\end{figure}

\begin{figure}
    \centering
    \includegraphics[width=0.425\textwidth]{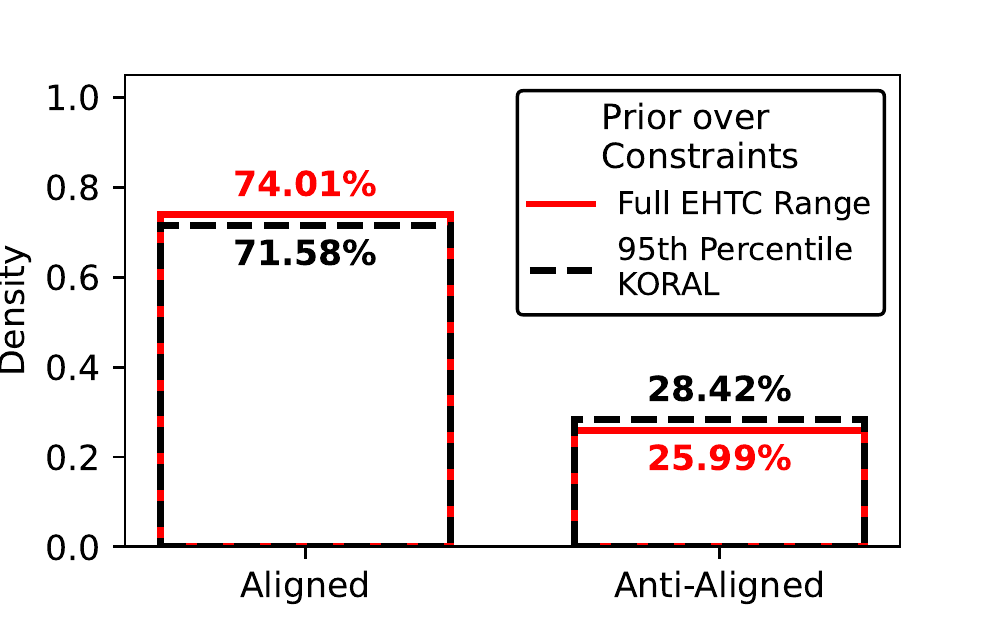}
    \caption{Categorical posterior distributions of magnetic field alignment relative to disk angular momentum from independent uniform priors on observed M87* constraints \citep{EHTC+2021a} (red) and further constrained to lie within the 95th percentile of our M87* image library (dashed black). Our models mildly prefer magnetic fields aligned with the disk angular momentum, but this is not very constrained in part because M87*'s circular polarization has not been detected. }
    \label{fig:m87_pred_bfield}
\end{figure}

We take allowable parameter ranges of various observables for M87* from \cite{EHTC+2021b} and \cite{EHTC+2019d} (for the image size), as listed in \autoref{tab:m87_params}, and evaluate them against our M87* image library. Of the 97,308 images in the library, a mere 154 images fall within the allowable constraints. Of the passing images, 133 are from retrograde models, 13 have $a_\bullet = 0$, and 8 are prograde models. All but 2 images have $R_\textrm{high} \geq 80$. Three models have more than 15 passing images. Two are consistent with EHTC Paper VIII: $a_\bullet = -0.5$ and $R_\textrm{high} = 160$ with aligned magnetic field and $a_\bullet = -0.7$, $R_\textrm{high} = 160$ with aligned magnetic field. The third, with $a_\bullet = -0.3$ and $R_\textrm{high}=160$, has an accretion flow anti-aligned $B$-field and therefore falls outside of the scope of the EHTC Paper VIII analysis. 

\begin{table}
\centering 
\begin{tabular}{c|c}
Parameter & Allowable Range for M87* \\
\hline
$|m|_\mathrm{net}$ & [0.01, 0.037] \\
$v_\mathrm{net}$ & [-0.008, 0.008] \\
$A$ & [2.0, 2.9] \\
$|\beta_2|$ & [0.04, 0.07] \\
$\angle\beta_2$ & [$-163^\circ$, $-127^\circ$] \\
Size & [$38 ~\mu\mathrm{as}$, $78 ~\mu\mathrm{as}$]
\end{tabular}
\caption{Allowable parameter ranges for various observables for M87* based on \citet{EHTC+2021a} and \citet{EHTC+2019d}.}
\label{tab:m87_params}
\end{table}

We also use the allowable parameter ranges as inputs to our random forest models and generate posterior distributions of the spin, $R_\textrm{high}$, and magnetic field polarity of M87*. We assume a uniform distribution as a prior over the allowable parameter ranges in each observable quantity. We further assume independence of each constraint and take our joint prior over all constraints to be independent in each quantity. We sample our joint prior to form a large test distribution of possible M87* observables. We train random forest models on our ray-traced library of M87* observables but in contrast to the models described earlier in \autoref{sec:rfs}, we exclude $\beta_0$, $\beta_1$, and $\beta_3$ information as EHT papers VII and VIII do not constrain these observables. 

After training, we ask our random forest models to predict spin, $R_\textrm{high}$, and the $B$-field alignment over each test distribution; these predictions are shown in Figures \ref{fig:m87_pred_spin}, \ref{fig:m87_pred_rhigh}, and \ref{fig:m87_pred_bfield}, respectively, in red. These predictive distributions show that we generally prefer high retrograde spins between $a_\bullet = -0.7$ and $a_\bullet = -0.4$ with some density around $a_\bullet = 0$. We also generally prefer large values of $R_\textrm{high}$, with some density near $R_\textrm{high} = 20$ and $R_\textrm{high} = 80$. Both of these distributions are generally consistent with the conclusions of \cite{EHTC+2021b}. Finally, looking at predictions of the $B$-field direction, our model weakly prefers accretion-aligned $B$-fields with $\sim 74\%$ of predictions suggesting an aligned $B$-field. We also note that though we present results starting from uniform priors over the allowable parameter space, we have found that the choice of prior does not substantially affect the predicted distribution of each quantity.

We find that our posterior distributions exhibit strange detailed structure, and we track most of this to library incompleteness related to the image size constraint.  The \cite{EHTC+2019d} constraint on image size is wide and includes smaller and larger image sizes than we observe in our own ray-traced M87* library. Thus, when performing inference in that area of parameter space, our random forest model extrapolates out-of-distribution. Considering this, we construct a second, more restrictive prior from the original EHT constraint ranges by additionally requiring that the observable values also lie within the 95th percentiles of our KORAL library. Essentially, this imposes a prior that the real system must lie within our image library. This, for example, restricts the values in our prior over image size to lie between roughly $[50.7 ~\mu\mathrm{as}, 65.2 ~\mu\mathrm{as}]$, 64 per cent more narrow than the full \cite{EHTC+2019d} range. The 95th percentile restriction for $|\beta_2|$ is 11 per cent more restrictive than the full \cite{EHTC+2021a} range. For all observables other than size and $|\beta_2|$, this stronger restriction eliminates a negligible portion of the original constraint range, meaning our library spans a wider range of observable values than existing EHT constraints. 

We perform inference on this restricted prior and show results for spin, $R_\mathrm{high}$, and $B$-field alignment in Figures \ref{fig:m87_pred_spin}, \ref{fig:m87_pred_rhigh}, and \ref{fig:m87_pred_bfield}, respectively, in black. We find that our inferred posterior for spin shifts density from near $a_\bullet = -0.7$ to $a_\bullet = -0.5$. For predicting $R_\mathrm{high}$, our restricted prior eliminates nearly all of the posterior density for $R_\mathrm{high} < 80$. Finally, inference of the $B$-field polarity is not substantially affected. This analysis underscores the incompleteness of our library. In particular, models including non-thermal electrons can produce larger images \citep{Ozel+2000,Mao+2017}.

\begin{figure}
    \centering
    \includegraphics[width=0.45\textwidth]{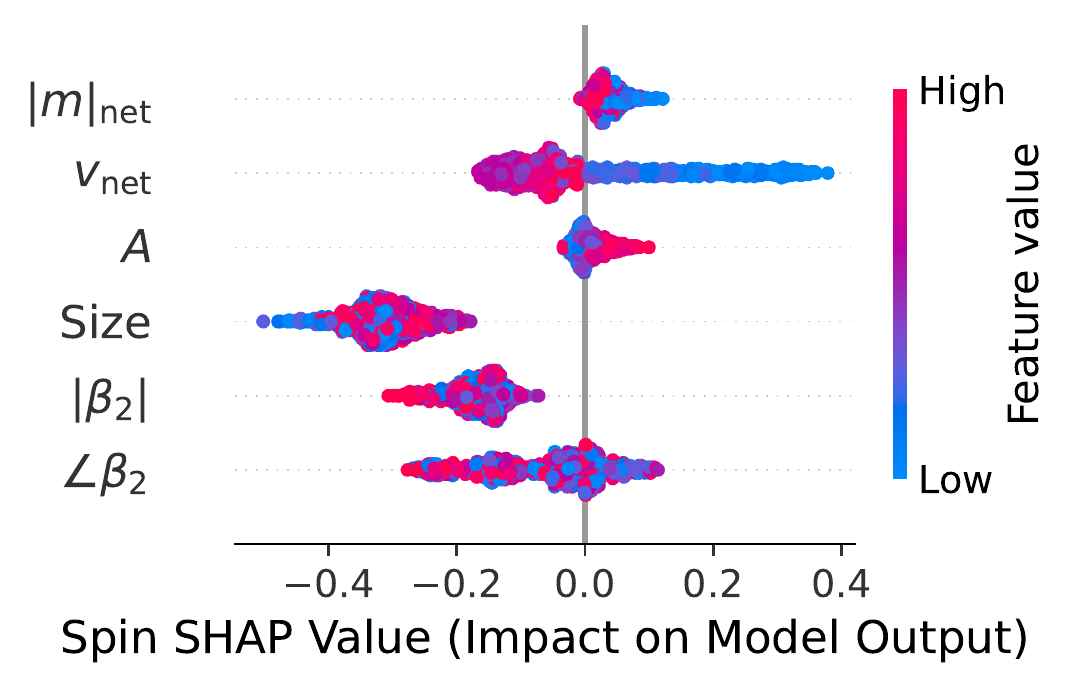}
    \caption{For each image feature, we visualize the distribution of its SHAP values for predicting the spin over our sampled M87* constraint distribution. As in \autoref{fig:sgrA_spin_shap_summary}, the colour encodes the relative value of each feature over the full range of values that feature spans. 
    \label{fig:m87_spin_shap_summary}}
\end{figure}

Finally, we examine the spin SHAP values from predictions on our full EHTC constraint prior. Though this is always the case, we make particular note here that SHAP values are calculated relative to the training distribution (our M87* image library) rather than the testing distribution (the EHTC constraint prior). This means, for example, that the sum of all spin SHAP values across the testing distribution may not average to 0. Instead, they average to the mean model prediction in the posterior, which is negative. We show distributions of SHAP values for predictions over our EHTC prior in \autoref{fig:m87_spin_shap_summary}. We find that that in general, the image size and $\beta_2$ lead the model to predict large retrograde values for the spin. Constraints on $|m|_\mathrm{net}$ and image asymmetry have relatively moderate effect on our posteriors for spin. Finally, we observe that $v_\mathrm{net} < 0$ has positive SHAP values, corresponding to the posterior density near $a_\bullet = 0$. 

As noted earlier, the posteriors for spin and $R_\textrm{high}$ produced by our random forests are consistent with the analysis in \cite{EHTC+2021b}, which eliminates images which do not pass all observational constraints. Additionally, our approach has the advantage of generating continuous posterior distributions over parameters of interest, which have concrete widths and predictive uncertainties. Further, by analyzing our models using SHAP or other interpretability tools, we can investigate the statistical relations learned by our model between observables and BH and accretion parameters of interest.

\section{Discussion and Conclusions}
\label{sec:discussion}

We have generated a library of 535,194 images for both M87* and Sgr A* derived from 9 GRMHD simulations with strong magnetic fields (MAD regime of accretion).  We compute observable quantities from each image and train a random forest machine learning algorithm to infer spin, inclination, the ion-to-electron temperature ratio, and magnetic field polarity from these quantities.  Our results are summarised as follows:

\begin{itemize}
    \item In the context of a very large but nevertheless incomplete GRMHD library, we have shown that spatially resolved polarimetric observables that are currently accessible to the EHT, can be used to indirectly constrain spin, inclination and the ion-to-electron temperature ratio using a random forest algorithm.
    
    \item For predicting these parameters, spatially resolved linear polarization stands out as the most important type of observation.  In particular, the twisty morphology of the linear polarization pattern, encapsulated in the parameter $\beta_2$, stands out as the most important observable, consistent with trends found in previous studies \citep{Palumbo+2020,Emami+2022}. 
    
    \item Based on current EHT constraints of M87*, our modelling prefers retrograde accretion disks around a relatively rapidly spinning black hole, and significantly colder electrons than ions in large $\beta$ regions.
    
    \item We have demonstrated that repeated EHT/ngEHT observations substantially tightens constraints on spin.  This is because turbulence in the accretion flow causes each of the observables to fluctuate.
    
\end{itemize}

The main limitation of our methodology is the incompleteness of our simulation set.  We have only considered MAD simulations in our study, guided by the polarimetric study of M87* performed by \citet{EHTC+2021b}.  There are many ways in which the simulation library could be expanded, which are beyond the scope of a single paper:
\begin{itemize}
    \item We have only considered one value of $R_\mathrm{low}$, which may deviate from 1 in the case of M87*, where radiative cooling may be important.  Moreover, alternatives to the $R_\mathrm{high}$-$R_\mathrm{low}$ temperature prescription adopted here exist in the literature that we have not explored \citep[e.g.,][]{Anantua+2020}.  Different temperature prescriptions can change the impact of Faraday rotation as well as the location at which emission occurs.  Self-consistently including cooling in GRMHD simulations can also impact the structure of M87* models \citep[e.g.,][]{Chael+2019,Yoon+2020}, and the implications for polarized signatures have yet to be fully explored.
    
    \item We have only considered thermal electron distribution functions, whereas the spectrum of Sgr A* motivates a non-thermal electron distribution \citep{Ozel+2000}, also predicted by particle-in-cell simulations \citep{Ball+2018}.  Models with a high-energy tail of non-thermal electrons tend to produce larger images \citep{Ozel+2000,Mao+2017}, and their impact on polarimetry remains understudied. 
    
    \item Our models include only perfectly aligned or anti-aligned disks, while in general these sources may be fed from a tilted disk \citep[e.g.,][]{Fragile+2007,Liska+2021}. This may alter the structure of images at large image radius.
    
    \item Our models assume an electron-ion plasma, while electron-positron pairs may also form in these systems \citep{Wardle+1998,Moscibrodzka+2011,Broderick&Tchekhovskoy2015,Wong+2021}.  The most important difference is that neither Faraday rotation nor emission of circularly polarization occur in pair plasmas, which may have a strong impact on all of the polarimetric observables considered here if pairs are produced in substantial quantities.  

    \item We have only considered observations at 230 GHz. However, in the near future, the EHT will produce images at 345 GHz, and the ngEHT will observe at 86 GHz. Since the magnetic field geometry drives the linear polarization structure and MAD models are not very Faraday thick \citep{Emami+2022}, we do not expect dramatic changes to single-frequency metrics at 345 GHz compared to 230 GHz. Images at 86 GHz may show more changes, however, as models may transition to becoming more optically and/or Faraday thick. Depending on the sensitivity of observations, spectral index and rotation measures may provide additional insights and constraints on BH and magnetic field properties. For example, \cite{Ricarte+2023} show that spectral index is sensitive to inclination of Sgr A*, and \cite{Ricarte+2020} find that rotation measure can provide insights toward magnetic field geometry and the ``cold'' electron population.  In all, though multi-wavelength observations and constraints are beyond the scope of this work, they represent an interesting area for future work.
\end{itemize}

Our methodology is also limited in our approach to modelling and inference. We have only considered one particular class of machine learning algorithms, random forests, whose input we limit to pre-defined image observables. Though random forests typically have strong performance compared to other methods (see e.g. \cite{Caruna+2006} or \cite{borisov+2021}), other algorithms could nevertheless learn different trends within the data and achieve better predictive accuracy, particularly if they have access to a broader set of image information. Though beyond the scope of a this paper, deep neural networks in particular could learn much richer polarimetric image features by not only learning trends between the input data, but by learning the features themselves, potentially discover new polarimetric observables the EHT can target. 

For simplicity, we have trained independent models for predicting each property of interest. A combined model which simultaneously predicts all properties at once can benefit from explicit mutual information (for example, when training to predict spin, such a model would not need to implicitly infer the inclination as an intermediate step) and achieve improved performance over the models we have presented in this work. However, such a model would be more complex to train and interpret, and we leave this investigation to future studies. 

Finally, our posterior distributions are generated by sampling uncertainty bounds on EHT observations. Though beyond the scope of this paper, Bayesian models can generate posteriors by directly and explicitly incorporating distributional uncertainty bounds from observations. Future studies exploring Bayesian models to perform inference with polarimetric data may be interesting to explore.

EHT polarimetric imaging has enabled new capabilities for black hole accretion flow science on event horizon scales.  The theoretical interpretation of these images involves an enormous modelling space, as detailed aspects of gravity, magnetohydrodynamics, and plasma physics all play a role.  Machine learning is ideal for bridging the gap between these two rich datasets.

In the future, EHT will continue to observe Sgr A* and M87*, which as we have shown, will help reduce uncertainties in our inferences related to the time variability of our sources.  The ngEHT will produce much more detailed maps with orders of magnitude more dynamic range and a greater field of view that will enable movies of both disk and jet.  For these new datasets, additional observable metrics will need to be devised that will be sensitive to fainter and more detailed features.  For this, parallel explorations with neural networks synergize well with our approach.  In general, the small hand-picked set of observables we consider in our study most likely misses informative aspects of our images that we have not noticed. It is possible that there are other, more informative observables that could further constrain our predictive distributions. Notably, we have also not considered the time variability and the frequency dependence of our models, which we expect could provide a wealth of new constraining power.  With the inclusion of more observational data and improved modelling, the performance of our random forest model is likely a lower limit to how tightly we can constrain the properties of the largest SMBHs on the sky.

\section{Acknowledgements}

We thank Ben~S. Prather for his thoughtful comments that improved the quality of this paper. RQ, AR and RN acknowledge support by the National Science Foundation under Grant No. OISE 1743747, as well as the support of grants from the Gordon and Betty Moore Foundation and the John Templeton Foundation. The opinions expressed in this publication are those of the author(s) and do not necessarily reflect the views of the Moore or Templeton Foundations. GNW gratefully acknowledges support from the Taplin Fellowship.

\section{Data Availability}

The data underlying the figures in this article will be shared on reasonable request to the corresponding author.

\bibliography{ms}

\appendix
\label{sec:appendix}

\section{Long Time Evolution of Simulations} \label{sec:time_evolution}

\begin{figure}
  \centering
  \includegraphics[width=0.45\textwidth]{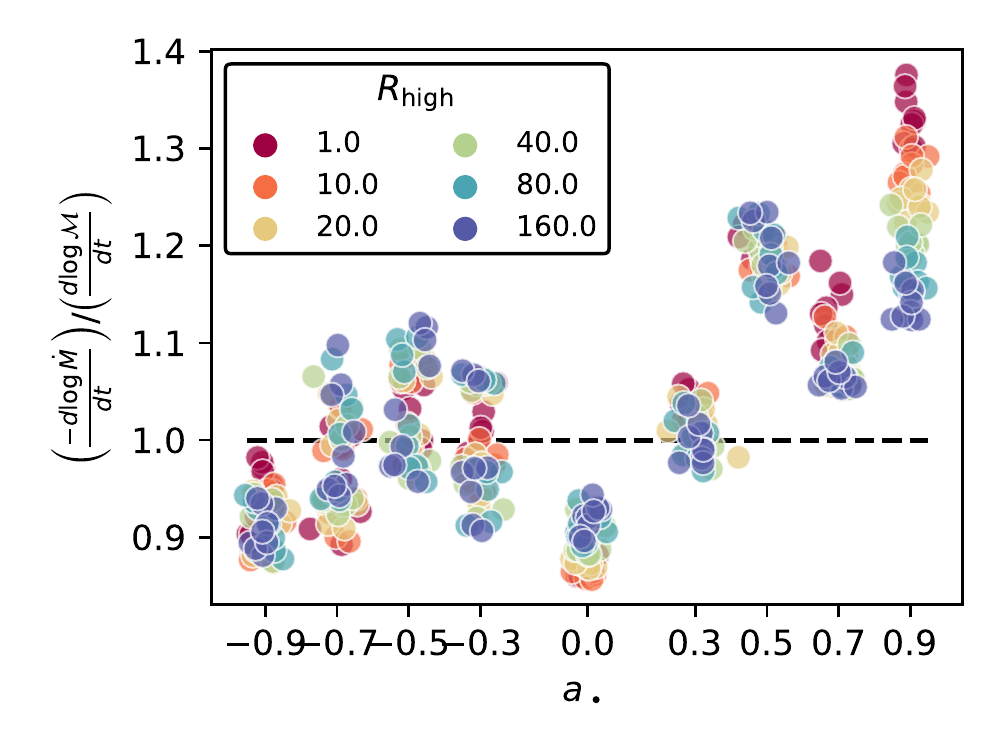}
  \caption{Comparison of log-derivatives of $\mathcal{M}$, fit directly from imaging, and mass accretion rate, obtained directly from GRMHD simulations. Each point represents a distinct time series with fixed spin, $R_\textrm{high}$, and observer inclination. A small amount of horizontal jitter has been added purely for visualisation purposes. 
  \label{fig:rapidity}}
\end{figure}

One might expect that the scaling in $\mathcal{M}$ should be inversely proportional to the decrease in $\dot{M}$ in the simulations. We examine this possibility in Figure \ref{fig:rapidity} by comparing our intensity-fitted $\mathcal{M}$ scaling to the direct logarithmic decrease in $\dot{M}$. The scaling in $\dot{M}$ is a reasonable approximation to $\mathcal{M}$ to first order, but does not capture all of the variation in $\mathcal{M}$ scaling. This suggests there are more effects than just $\dot{M}$ scaling at play. In particular, there may be fluctuations in temperature and scaling in $B^2$ which are captured by $\mathcal{M}$ but not $\dot{M}$. Additionally, $\dot{M}$ may not trace the density in the emitting region perfectly. 

\begin{figure*}
  \centering
  \includegraphics[width=0.9\textwidth]{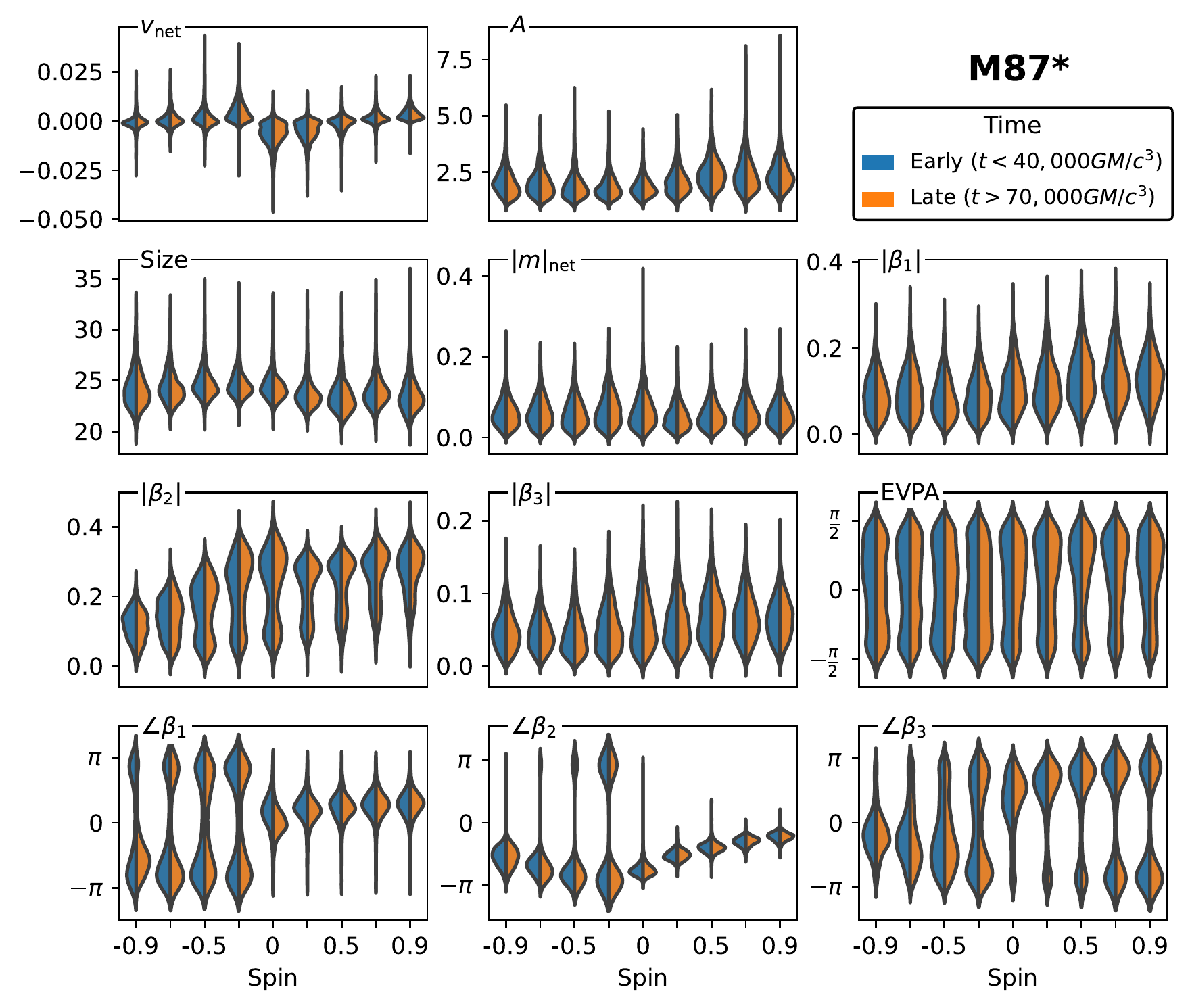}
  \caption{Distributions of each observable quantity for our M87* library, partitioned into early snapshots from the first third of each image sequence ($10,000~GM/c^3 < t < 40,000~GM/c^3$) and late snapshots from the final third ($70,000~GM/c^3 < t < 100,000~GM/c^3$). We note that because of the finite width kernel used to compute the densities, the violin plots may extend beyond the true range of allowable values (e.g. $|\angle\beta_j| > \pi$).   We observe no significant time evolution in the distributions of any observable quantity. 
  \label{fig:time_stability}}
\end{figure*}

One of the primary motivations for generating this set of long GRMHD simulations was to test whether any of the observables change as these simulations evolve over long timescales.  Apart from the expected decrease in the overall accretion rate associated with the draining and relaxation of the torus, which we remove when imaging, we do not find any evidence of significant time evolution.  We show distributions of observables for early and late snapshots in our M87* library in \autoref{fig:time_stability}; similarly we do not find any evidence of time evolution within our resolved image observables. This bolsters our confidence in results from simulations run for much shorter periods of time.

\section{Comparison to EHT GRMHD Library}

As a validation test, we perform spin inference on an image library ray-traced from an independent set of GRMHD simulations. In particular, we use a subset of the M87* library first presented in \citet{EHTC+2021b}. This validation image library contains 1230 images at 5 spins ($a_\bullet = \{-0.94, -0.5, 0, 0.5, 0.94\}$) and 6 values of $R_\mathrm{high}$ ($1$, $10$, $20$, $40$, $80$, $160$). As with our KORAL image library, we flip the viewing inclination such that the jet direction is consistent with observations of the orientation of the jet. In particular, prograde and 0 spin models are ray traced at an inclination of $i = 163^\circ$ and retrograde models with inclination of $i = 17^\circ$. 

\begin{figure}
  \centering
  \includegraphics[width=0.45\textwidth]{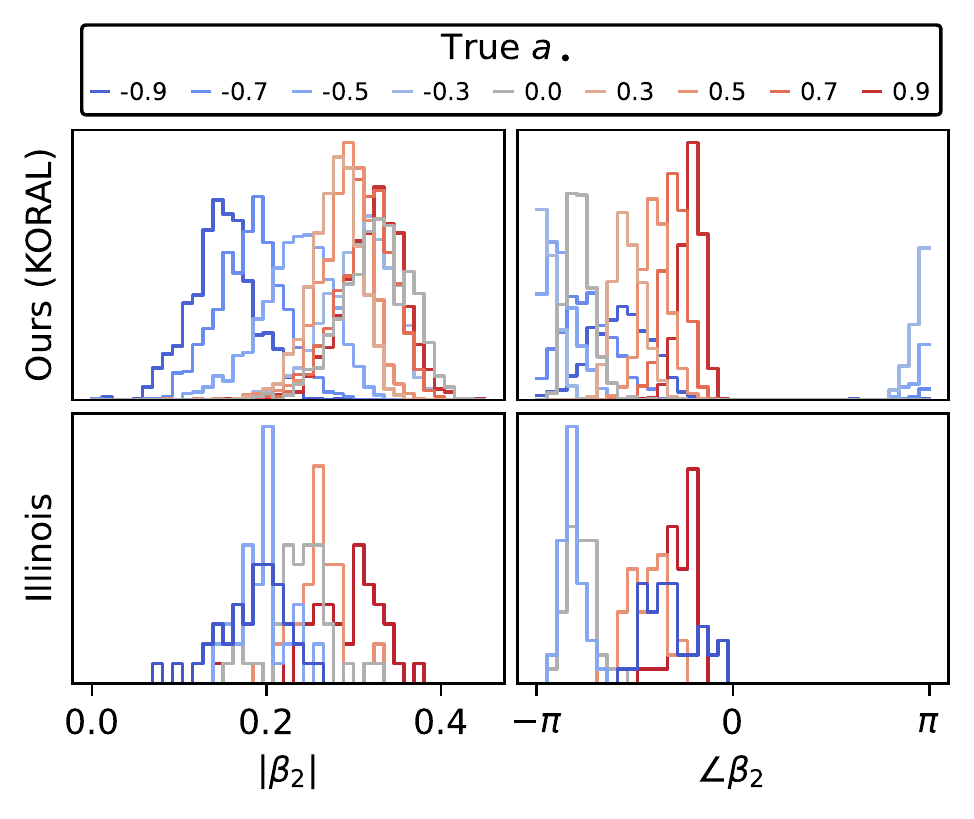}
  \caption{Distributions of $\beta_2$ for our KORAL image library and the Illinois image library across various values of spin and $R_\mathrm{high} = 20$. We map the colour of $a_\bullet = \pm 0.94$ to $a_\bullet = \pm 0.9$ for visual consistency. Though the libraries were generated with independent GRMHD models, the distributions are generally robust.  
  \label{fig:koral_il_beta_comparison}}
\end{figure}

We blur each image with a $20 \mu$as Gaussian beam and calculate the 11 observable image quantities described in \autoref{sec:observables}. As indicated by the feature importances shown in \autoref{fig:m87_feature_importances}, the most important feature for predicting the spin of M87* is $\beta_2$. Thus, we show a comparison of the distributions of $|\beta_2|
$ and $\angle\beta_2$ between our KORAL image library and the Illinois image library in \autoref{fig:koral_il_beta_comparison}. The distributions of $|\beta_2|$ and $\angle\beta_2$ are consistent between the two image libraries, despite being generated from different GRMHD models. Though not shown here, we observe similar consistency with the other computed image observables. This suggests that our chosen computed image observables are broadly consistent  between different GRMHD schemes. We note, however, that the initial conditions of these simulations were both magnetized \citet{Fishbone&Moncrief1976} tori in hydrostatic equilibrium with initially dipolar fields.  It will be an important next step to explore the relative consistency of simulations with different initial conditions.

\begin{figure}
  \centering
  \includegraphics[width=0.45\textwidth]{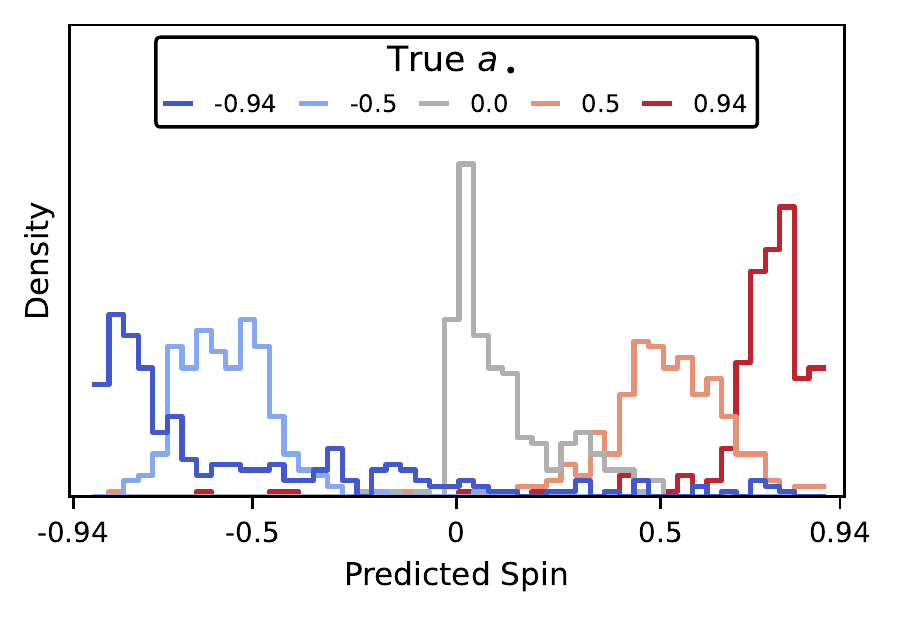}
  \caption{Predictions of spin on the Illinois v1 M87* library using a model trained on our KORAL M87* library. Despite being a different set of GRMHD models, our model is able to recover the true spin values with reasonable accuracy. We note that the model is limited to predicting values within its training data, i.e. $a_\bullet \in [-0.9, 0.9]$, but the extreme spins in the Illinois library are slightly beyond this range. 
  \label{fig:illinois_predictions}}
\end{figure}

Using our random forest model trained on our KORAL library to predict spin, we perform inference on the spin of each image in the Illinois library. The distributions of predicted spin values are shown in \autoref{fig:illinois_predictions}. We find that our model is able to recover the true spins with high accuracy. Each of the 5 spin values represented in the Illinois library is discriminated with little overlap in the predictive distributions. This suggests that our model is not overfitting to model-specific GRMHD and imaging parameters used to generate our KORAL image library. Instead, our model is learning more general underlying physical features robust to differences in GRMHD schemes. 

\section{Effect of Varying $R_\mathrm{low}$}
\label{sec:r_low_test}

We only consider $R_\mathrm{low} = 1$ models in our analysis due to computational expense, and we consider this to be an important limitation of our work.  Here, we briefly explore the effect of varying $R_\mathrm{low}$ to 10, a value that is also tested in previous EHT studies of M87* \citep[e.g.,][]{EHTC+2021b,Fromm+2022}.  This is motivated by the fact that radiative cooling may be significant enough in M87* to warrant additional suppression of the electron temperature in these models \citep[e.g.,][]{Moscibrodzka+2011,Ryan+2018,Chael+2019}.

\begin{figure}
  \centering
  \includegraphics[width=0.49\textwidth]{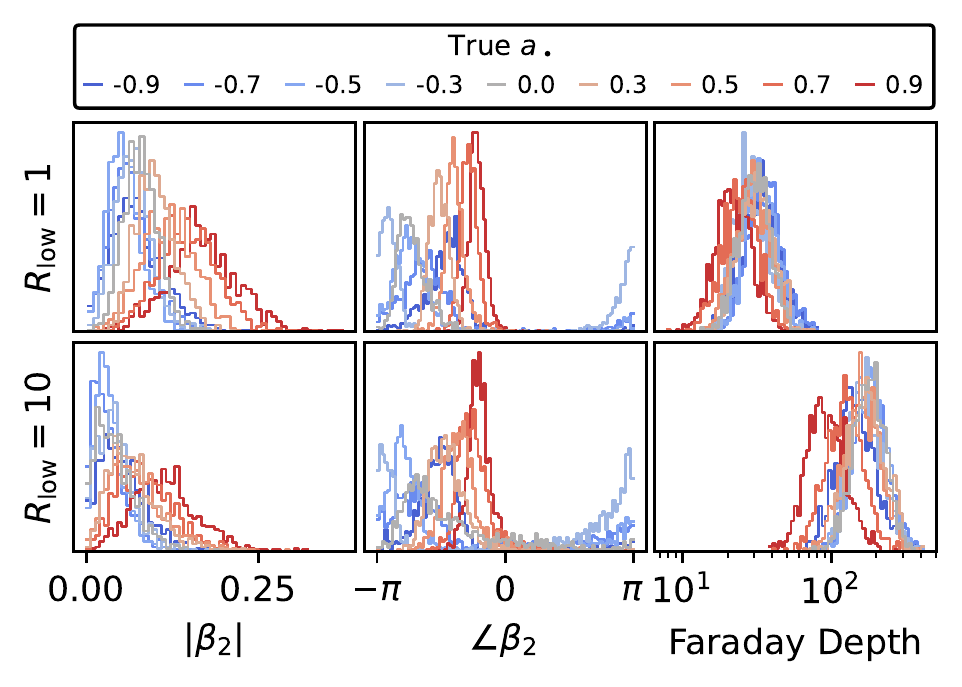}
  \caption{Distributions of $\beta_2$ and Faraday depth for the $R_\mathrm{low} = 1$ M87* library used in this paper and an $R_\mathrm{low} = 10$ M87* library. All distributions shown are restricted to $R_\mathrm{high} = 160$ and aligned magnetic fields. The distributions of $\angle \beta_2$ are generally robust while $|\beta_2|$ is moderately suppressed for $R_\mathrm{low} = 10$, due to the additional scrambling from increased Faraday depth.
  \label{fig:rlow10_comparison}}
\end{figure}

We recompute M87* images with $R_\mathrm{low}=10$ and all spins, but with fixed $R_\mathrm{high} = 160$ and aligned magnetic field.  We most carefully examine $\beta_2$, which is shown to be the most important observable in this work.  In \autoref{fig:rlow10_comparison}, we plot distributions of $|\beta_2|$, $\angle \beta_2$ for each of these $R_\mathrm{low}$ values as a function of spin.  As expected, increasing $R_\mathrm{low}$ (cooling the electrons) results in larger Faraday depths. 
This is because Faraday rotation is less efficient at higher temperatures, and these models need larger mass scalings in order to reproduce the 0.5 Jy core flux of M87*.  As a direct result, $|\beta_2|$ decreases as $R_\mathrm{low}$ increases.  Fortunately, although there are some small differences, the distributions of $\angle\beta_2$ are relatively robust.  This is consistent with \citet{Emami+2022}, who study the sensitivity of $\angle \beta_2$ to Faraday rotation in detail and generally find small shifts for MAD models.  We might expect larger differences for SANE models, which tend to be more Faraday thick \citep{Moscibrodzka+2016,Ricarte+2020,EHTC+2021b}.  Though not shown, we observe a similar degree of stability in other $\angle\beta_i$ modes and depolarization in $|\beta_i|$.  While some differences remain, it is also plausible that a random forest model trained on a complete library of both $R_\mathrm{low}=1$ and $R_\mathrm{low}=10$ models could learn to separate them about as well as it does $R_\mathrm{high}$ currently.

\end{document}